\documentclass[fleqn,12pt]{wlscirep}
\usepackage{setspace}
\usepackage{subcaption}
\usepackage{caption}
\usepackage{color}
\usepackage[symbol]{footmisc}
\usepackage{amsmath}
\usepackage{multirow}
\usepackage{adjustbox}
\usepackage{float}
\usepackage{threeparttable}
\usepackage{txfonts}
\usepackage{ab}
\newcommand{\EFig}{Extended Data Fig.}
\newcommand{\ETab}{Extended Data Tab.}

\title{A dynamic magneto-ionic environment around a long-period radio transient}

\author[1] {Botao Li}
\author[1] {Yongjie Jin}
\author[1] {Shike Qu}
\author[1] {Chaohua Gao}
\author[1] {Wei Wang}
\author[2] {Pei Wang}
\author[3] {Shumei Jia}
\author[2] {Jifeng Liu}

\affil[1]{Department of Astronomy, School of Physics and Technology, Wuhan University, Wuhan 430072, People's Republic of China;}
\affil[2]{National Astronomical Observatories, Chinese Academy of Sciences, Beijing 100012, People's Republic of China;}
\affil[3]{Key Laboratory of Particle Astrophysics, Institute of High Energy Physics, Chinese Academy of Sciences, Beijing 100049, People's Republic of China}
\begin{abstract}
\bf{
Faraday rotation provides one of the most direct probes of magnetized plasma along the line of sight, from the large-scale Galactic magnetic field \cite{1994han,2006han} to compact plasma environments local to the radio sources \cite{2022wangbebinary}. In particular, temporal variations in rotation measure can trace the variations of the density, magnetic-field strength, or geometry of the Faraday-rotating medium, e.g., radio pulsars \cite{2023galacticcenter} and fast radio bursts \cite{2022wangbebinary,2025Sfastrepeaterrm}. Long-period radio transients (LPTs), a recently identified class of coherent radio sources with periods of hundreds to thousands of seconds \cite{2022Nat76s,dong2025chime,2022Nat18min,li202444,wang2025detection,rea2026long,zelati2024ultra,hurley2023long,caleb2024emission,men2025highly,lee2025emission}, remain poorly explored because their physical nature is still uncertain. Here we present full-Stokes observations of a 421-s LPT source CHIME J0630$+$25, and report, for the first time, the extreme rotation measure (RM) variability from approximately $-400$ to $-1630~{\rm rad~m^{-2}}$. The rapid RM changes occur on short timescales, including a difference of $\sim50~{\rm rad~m^{-2}}$ between two consecutive rotation periods and a sudden change of $\sim536~{\rm rad~m^{-2}}$ within $\sim 2500$ s. Compared with other radio-emitting sources, CHIME J0630$+$25 occupies an unusual region of the DM--RM plane: despite its very small ${\rm DM}\simeq22~{\rm pc~cm^{-3}}$, the $|{\rm RM}|$ reaches over $1.6\times10^{3}~{\rm rad~m^{-2}}$, far exceeding that expected from the interstellar medium and pulsars. The fast and large-amplitude RM variations should arise from a compact, structured, and rapidly evolving magneto-ionic environment local to the source, possibly associated with a compact binary system. 

}
\end{abstract}

\begin{document}

\maketitle


CHIME J0630$+$25 is a nearby long-period radio transient (LPT) with a period of $\sim421$ s and a small dispersion measure, ${\rm DM}\sim22~{\rm pc~cm^{-3}}$, corresponding to an estimated distance of $170^{+310}_{-100}$ pc \cite{dong2025chime}. Despite this small DM, CHIME J0630$+$25 was reported to have a large rotation measure, ${\rm RM}=-347.8\pm 0.6~{\rm rad~m^{-2}}$, suggesting that a substantial fraction of the Faraday rotation may arise from a magnetized environment local to the source. Its timing behaviour was described by a glitch-like model that suggests an isolated neutron star with a surface magnetic field of $\sim 1.5\times10^{15}$ G.

\begin{figure*}[htbp]
\centering

\begin{subfigure}{0.98\textwidth}
    \centering
    \includegraphics[width=\linewidth]{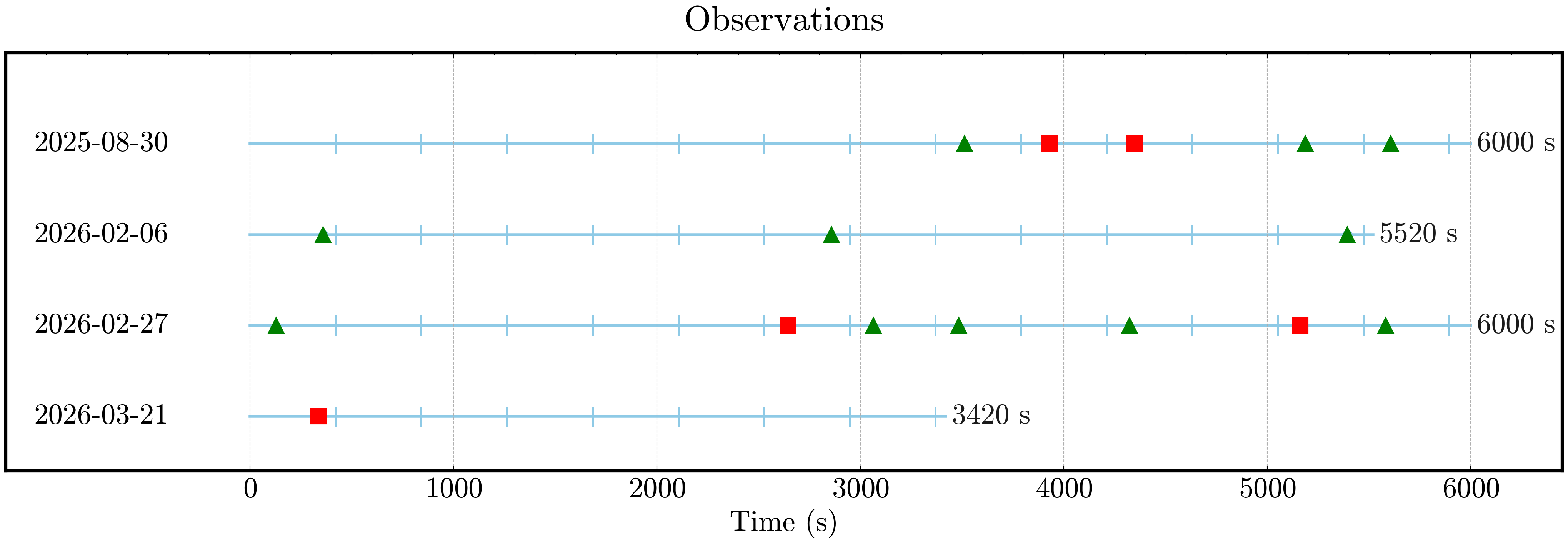}
    \caption{}
\end{subfigure}

\vspace{0.35cm}

\begin{subfigure}{0.32\textwidth}
    \centering
    \includegraphics[width=\linewidth]{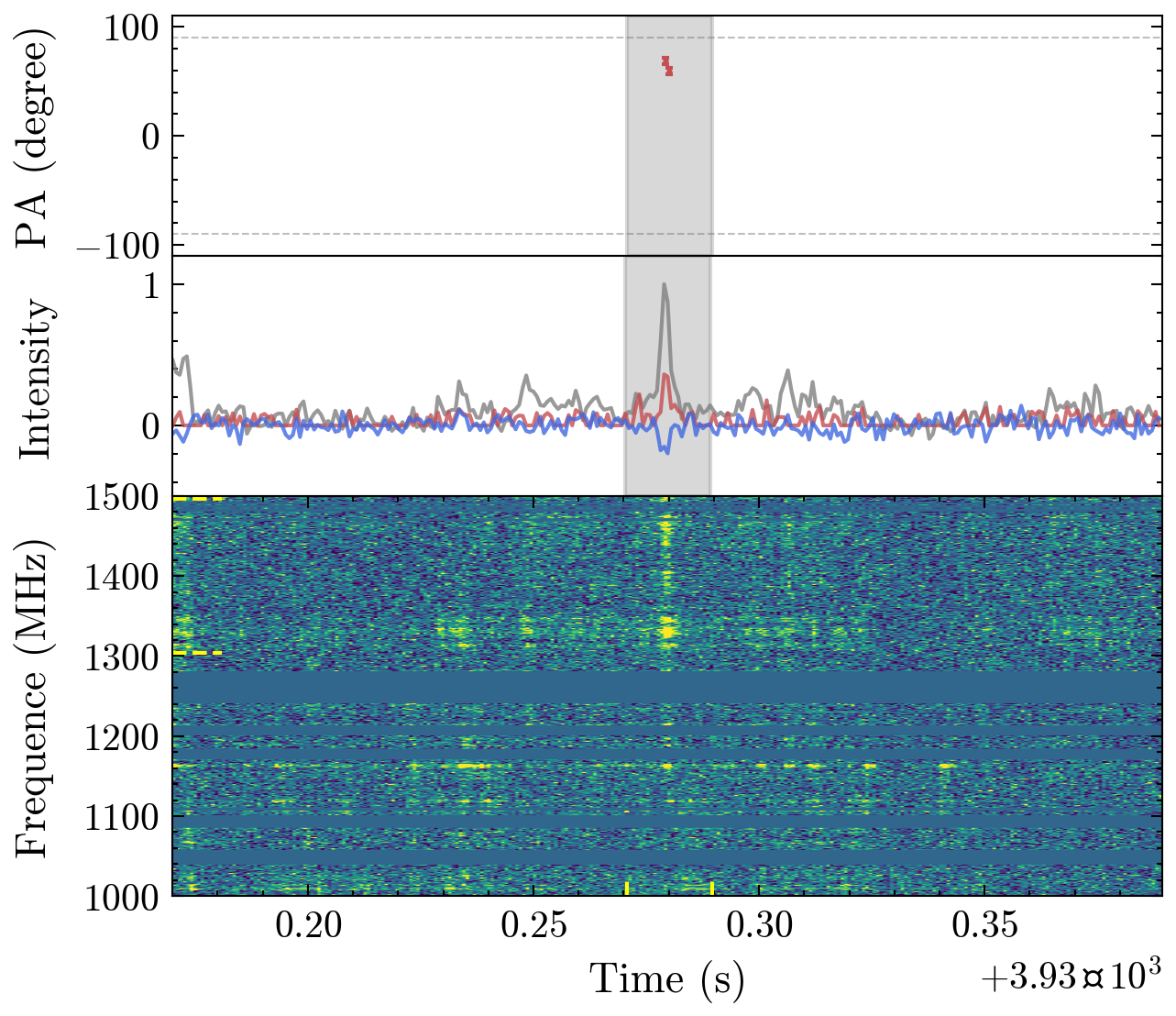}
    \caption{}
\end{subfigure}
\hfill
\begin{subfigure}{0.32\textwidth}
    \centering
    \includegraphics[width=\linewidth]{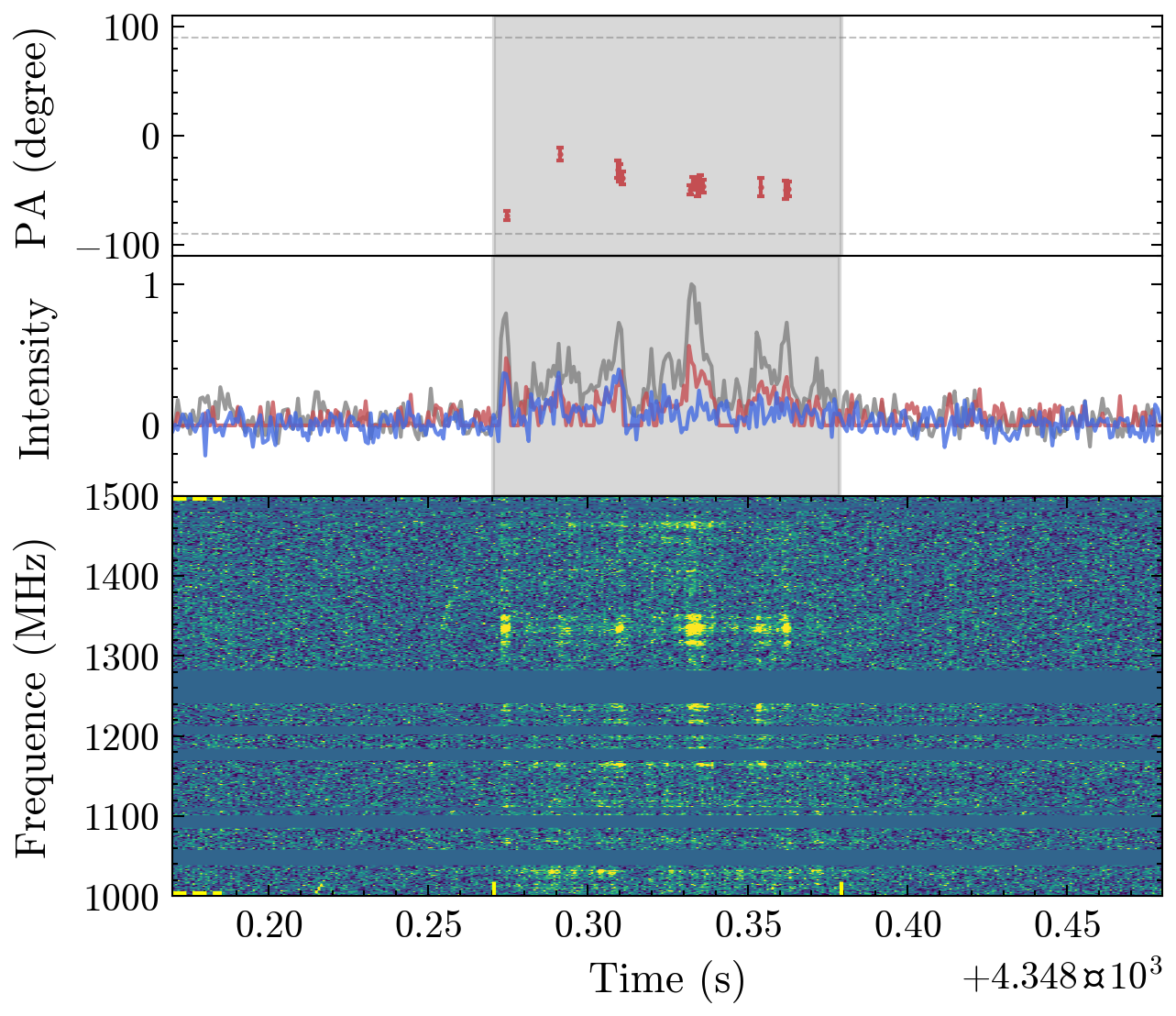}
    \caption{}
\end{subfigure}
\hfill
\begin{subfigure}{0.32\textwidth}
    \centering
    \includegraphics[width=\linewidth]{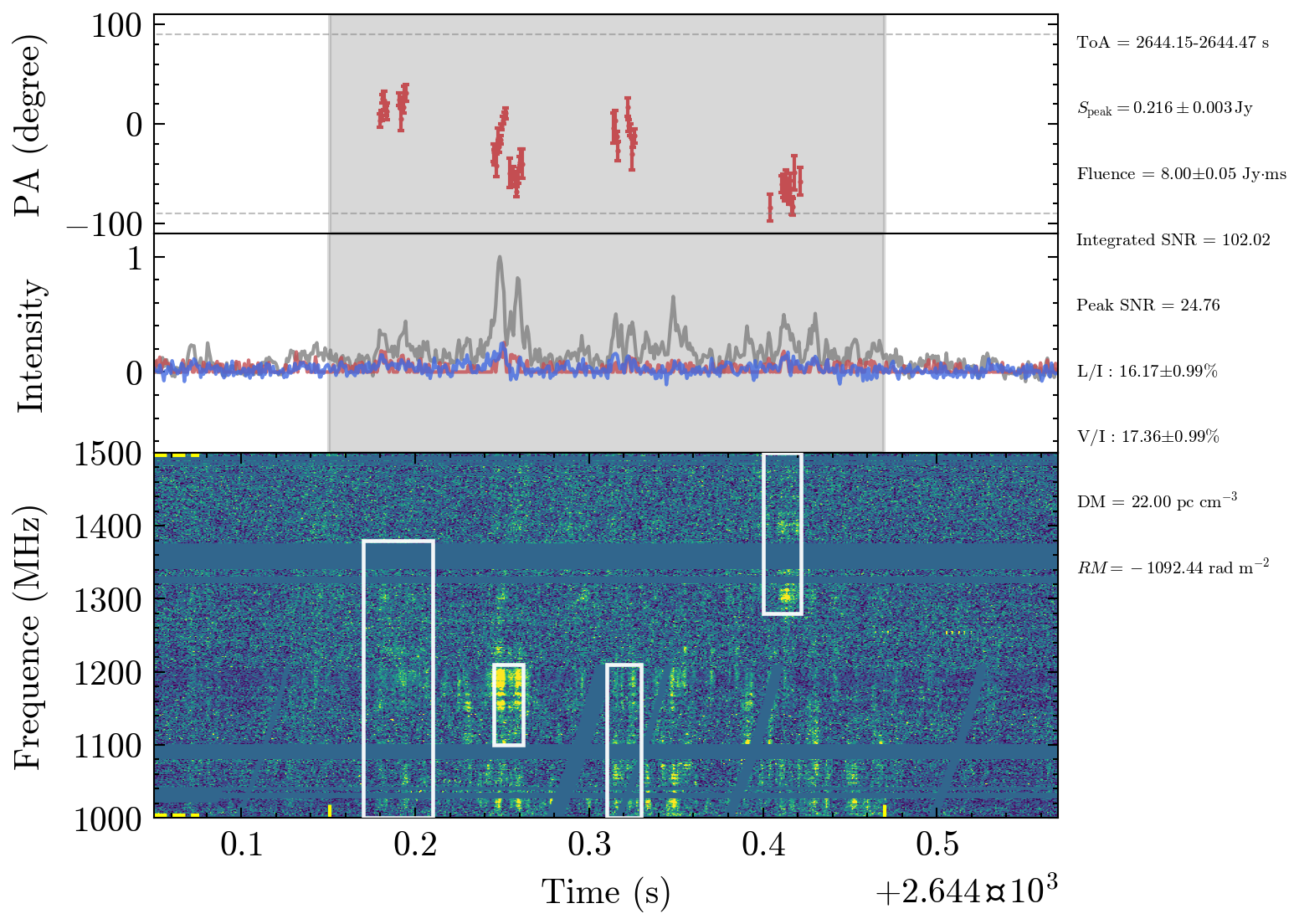}
    \caption{}
\end{subfigure}

\caption{
FAST observation information and representative bursts of CHIME J0630$+$25.
\textbf{(A)}: Summary of the four FAST observing sessions. Horizontal light-blue bars show the time span of each observation, labelled by date on the left and by total observing duration on the right. Short vertical ticks mark successive rotations using \(P\simeq421\) s. Green triangles mark detected radio bursts without reliable RM measurements, while red squares mark bursts with measurable RM. \textbf{(B)--(D)}, Three representative bursts. \textbf{(B)} and \textbf{(C)} show bursts observed on 2025-08-30, while \textbf{(D)} shows a burst observed on 2026-02-27. In each panel, the upper, middle and lower sub-panels show the PA, the light curves of Stokes \(I\), \(L\) and \(V\), and the dynamic spectrum of Stokes \(I\), respectively. The short yellow dashed lines indicate the time and frequency ranges used to generate the upper and middle sub-panels. In panel \textbf{(D)}, the white boxes mark the time-frequency windows used to calculate the PA curves. 
}
\label{fig:observations}
\end{figure*}

To investigate its local magneto-ionic environment, we carried out four full-Stokes observations of CHIME J0630$+$25 with the Five-hundred-meter Aperture Spherical Telescope (FAST) between 2025 August and 2026 March and detected pulses in all four epochs, confirming that the source remained active over a timescale of several months. Simultaneous \textit{Insight}-HXMT observations yielded no significant X-ray detection during the radio observing epochs, with the corresponding luminosity upper limit ($3\sigma$) of $\sim5\times10^{31}~{\rm erg~s^{-1}}$ from 2--100 keV. The observation summary and representative burst dynamic spectra are presented in Fig.~\ref{fig:observations} and \ETab~\ref{tab:obs}. The burst peak flux densities are typically only tens of mJy, exhibiting intermittent and weak linear polarization, generally below 50\%, while the strong circular polarization appears more commonly with its absolute fraction up to $20$--$70\%$ (see \ETab~\ref{tab:info} in Methods). The burst dynamic spectra show diverse morphologies, including strong frequency-dependent intensity modulation, and both upward- and downward-drifting features. The times of arrival (TOAs) in this paper are referenced to the highest observing frequency (1500 MHz), with the zero point defined at the beginning of each observation.

\begin{figure*}[htbp]
\centering

\includegraphics[width=0.95\textwidth]{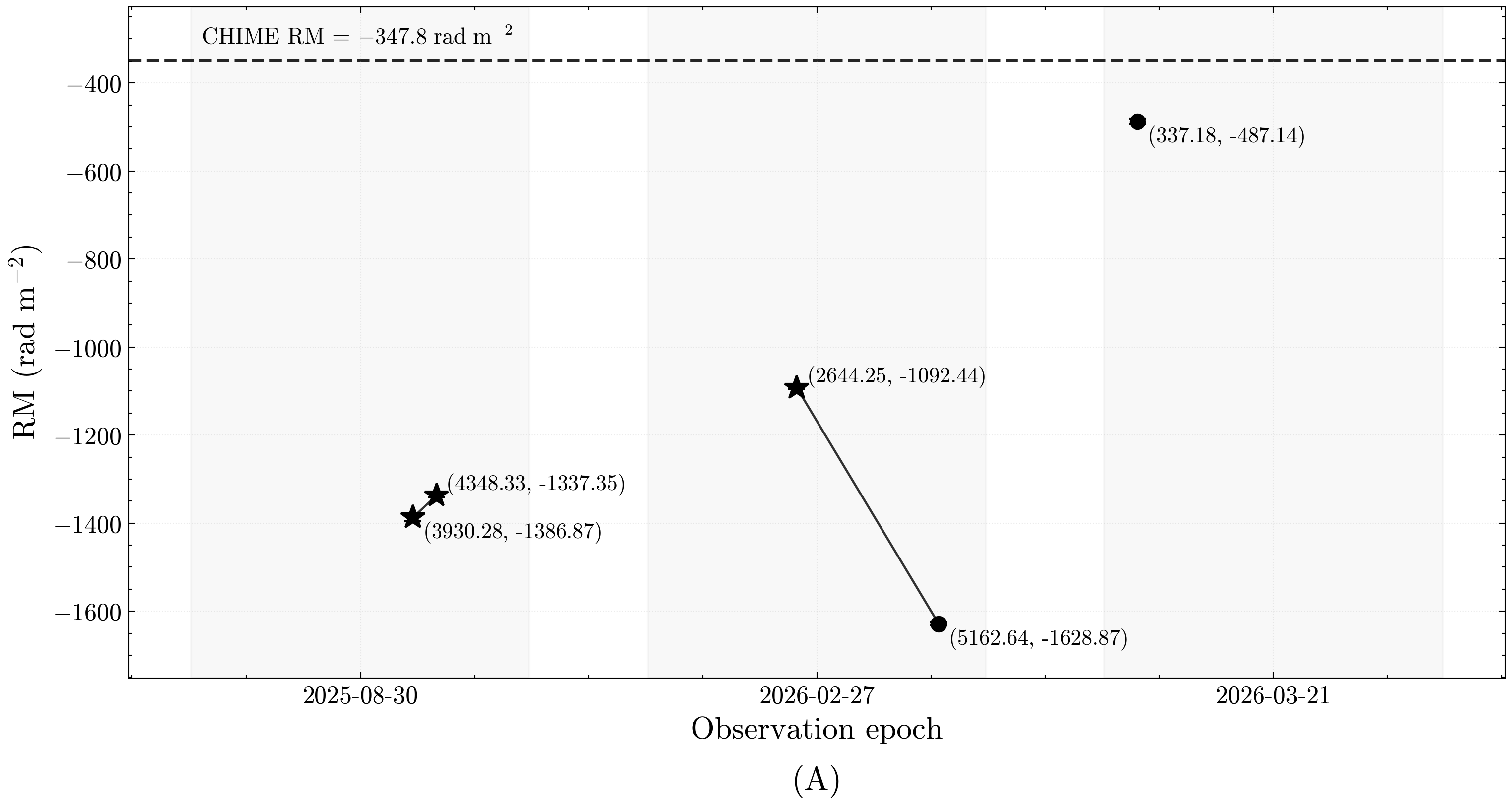}

\vspace{0.35cm}

\includegraphics[width=0.95\textwidth]{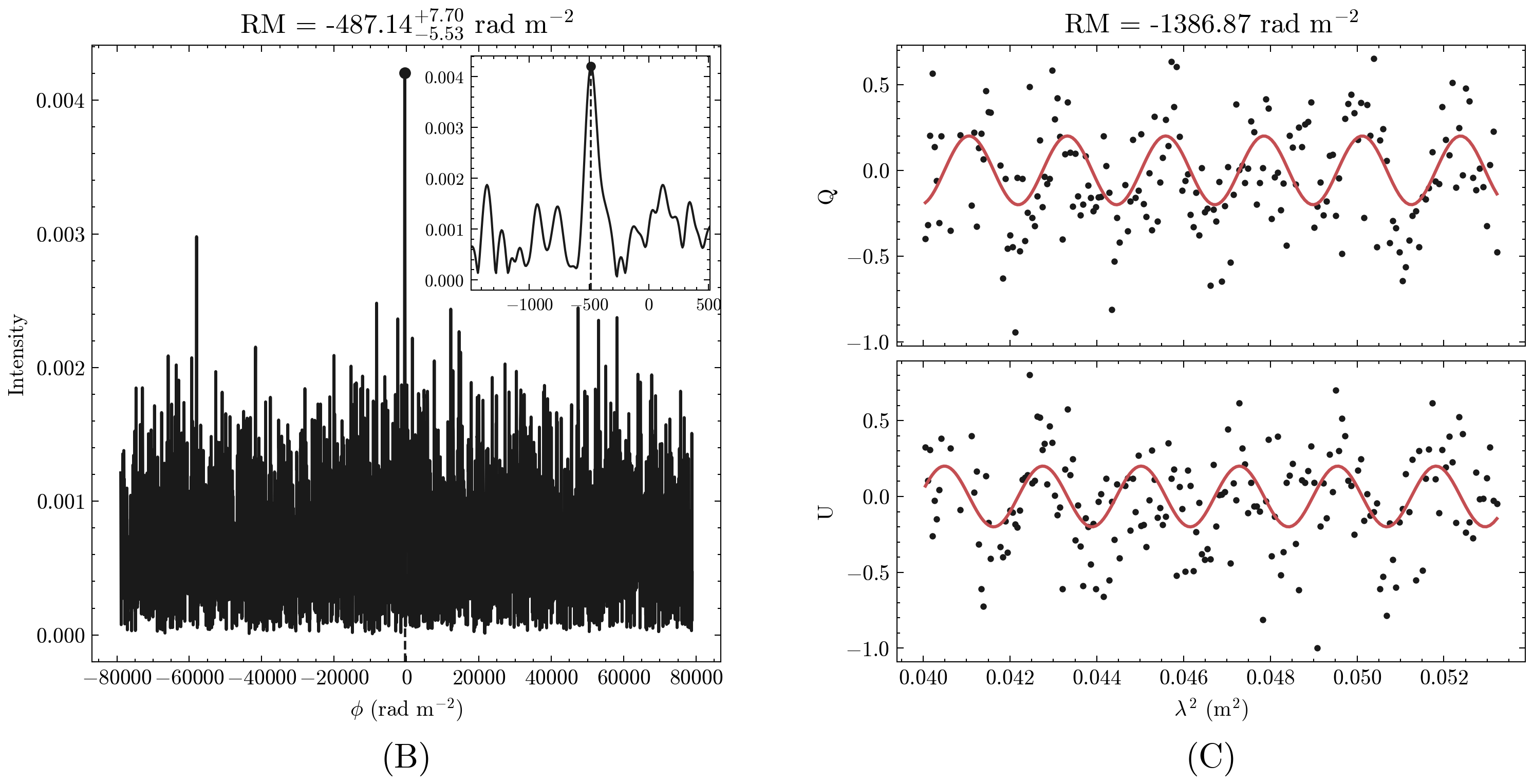}

\caption{
Rotation-measure evolution and representative RM measurements of CHIME J0630$+$25.
\textbf{(A)}: RM evolution measured with FAST with the TOAs noted for each measured RM value, together with the previously reported CHIME RM value. The star symbols indicate RM measurements for which the frequency-dependent Stokes $Q$ and $U$ spectra can be directly fitted, whereas circular symbols indicate RM values obtained from RM synthesis only.
\textbf{(B)}: RM-synthesis result for the 2026-03-21 burst, which gives an RM of $-487.14~{\rm rad~m^{-2}}$. \textbf{(C)}: Representative Stokes $Q/U$ spectra fitted using RM-synthesis result for the bright burst near $t\simeq3930$ s on 2025-08-30.
}
\label{fig:rmcal}
\end{figure*}

The most prominent result of our observations is that the RM of CHIME J0630$+$25 is extremely variable on very short timescales. In the brightest burst detected on 2025-08-30, clear oscillatory structures are visible in the Stokes $Q$ and $U$ spectra. RM synthesis and $Q/U$ fitting give ${\rm RM}=-1386.87^{+9.72}_{-10.15}$, substantially larger in magnitude than the value previously reported by CHIME. A pulse in the next burst cycle yields ${\rm RM}=-1337.35^{+3.72}_{-3.73}~{\rm rad~m^{-2}}$, showing a burst-to-burst RM difference of $\sim50~{\rm rad~m^{-2}}$. This change occurs over one rotation cycle and provides direct evidence that the Faraday-rotating medium toward the source varies on short timescales. The RM value continued to vary in later observations. On 2026-02-27, a linearly polarized pulse at $t\simeq2644$ s gives ${\rm RM}=-1092.44^{+3.04}_{-3.37}~{\rm rad~m^{-2}}$. About 2500 s later, after six burst cycles, another linearly polarized pulse gives ${\rm RM}=-1628.87^{+3.64}_{-2.86}~{\rm rad~m^{-2}}$. Thus, within a single observing session, the measured RM changed by $\sim536~{\rm rad~m^{-2}}$. On 2026-03-21, we measured ${\rm RM}=-487.14^{+7.70}_{-5.53}~{\rm rad~m^{-2}}$, closer to the original CHIME value but far from the 2025 August and 2026 February measurements. The RM evolution curve along with RM measurement examples is shown in Fig.~\ref{fig:rmcal}.


The ordinary interstellar medium is unlikely to account for the extreme and fast varying RM. A simple foreground estimate gives ${|\rm RM|}\sim {{\rm DM}\,|B_{\parallel}|\over 1.232}\simeq18~{\rm rad~m^{-2}}$ for ${\rm DM}=22~{\rm pc~cm^{-3}}$ and an average line-of-sight magnetic field of $|B_{\parallel}|=1~\mu{\rm G}$. The Galactic RM contribution inferred from all-sky Faraday maps is also expected to be only of order tens of ${\rm rad~m^{-2}}$ along this line of sight \cite{hutschenreuter2022galactic,dong2025chime}. The observed RM amplitude and its variation by more than ${1000~\rm rad~m^{-2}}$ therefore require an additional Faraday-rotating region local to the source. 

Beyond CHIME J0630$+$25, diverse RM behaviours have been observed in different radio-emitting sources, including ordinary pulsars, radio-emitting magnetars, binary pulsars and repeating FRBs. For most radio pulsars, RM is stable on long timescales \cite{2024meerkat}. Long-term monitoring of 20 millisecond pulsars, for example, found typical values of $|d{\rm RM}/dt| < 1~{\rm rad~m^{-2}~yr^{-1}}$ \cite{2011mspsr}. Radio-emitting magnetars \cite{2021swift,2012psr,2018psr,2020sgr1,2020sgr3,2021sgr2}, with the notable exception of the Galactic-centre magnetar PSR J1745$-$2900, also generally show relatively stable RM, including XTE J1810$-$197 \cite{2006XTE,2007XTE,2019XTE} and 1E 1547.0$-$5408 \cite{20081e,20231e}. Some pulsars exhibit phase-resolved RM variations \cite{2009phase,2019phase}, which are longitude-dependent apparent RM gradients across the pulse profile and are commonly attributed to magnetospheric or propagation effects rather than secular changes of an external Faraday screen.

Significant RM changes typically occur when the line of sight intersects dense and strongly magnetized plasma. Galactic-centre pulsars show measurable RM variability due to a turbulent and magnetized foreground \cite{2023galacticcenter}. The Galactic-centre magnetar PSR J1745$-$2900 displayed large RM variations of $\sim4000~{\rm rad~m^{-2}~yr^{-1}}$ over $\sim1500$ d while exhibiting relatively small DM changes, indicating an extremely magnetized environment \cite{2018gcp}. 
Time-variable RM is also observed in some repeating FRBs. FRB 121102 has an extremely large and variable RM, decreasing from $1.46\times10^{5}$ to $1.33\times10^{5}~{\rm rad~m^{-2}}$ over months, with day-scale fluctuations of $\sim200~{\rm rad~m^{-2}}$ \cite{2021frb121102}. FRB 20201124A exhibits long-term RM variations of $\sim500~{\rm rad~m^{-2}}$, which have been interpreted in the context of a magnetar/Be-star binary system \cite{2022wangbebinary}. However, the RM evolution in these repeating FRBs usually occurs on timescales of days to months, much longer than the minute-to-hour timescales observed in CHIME J0630$+$25.

\begin{figure*}[htbp]
\centering

\begin{subfigure}{0.7\textwidth}
    \centering
    \includegraphics[width=\linewidth]{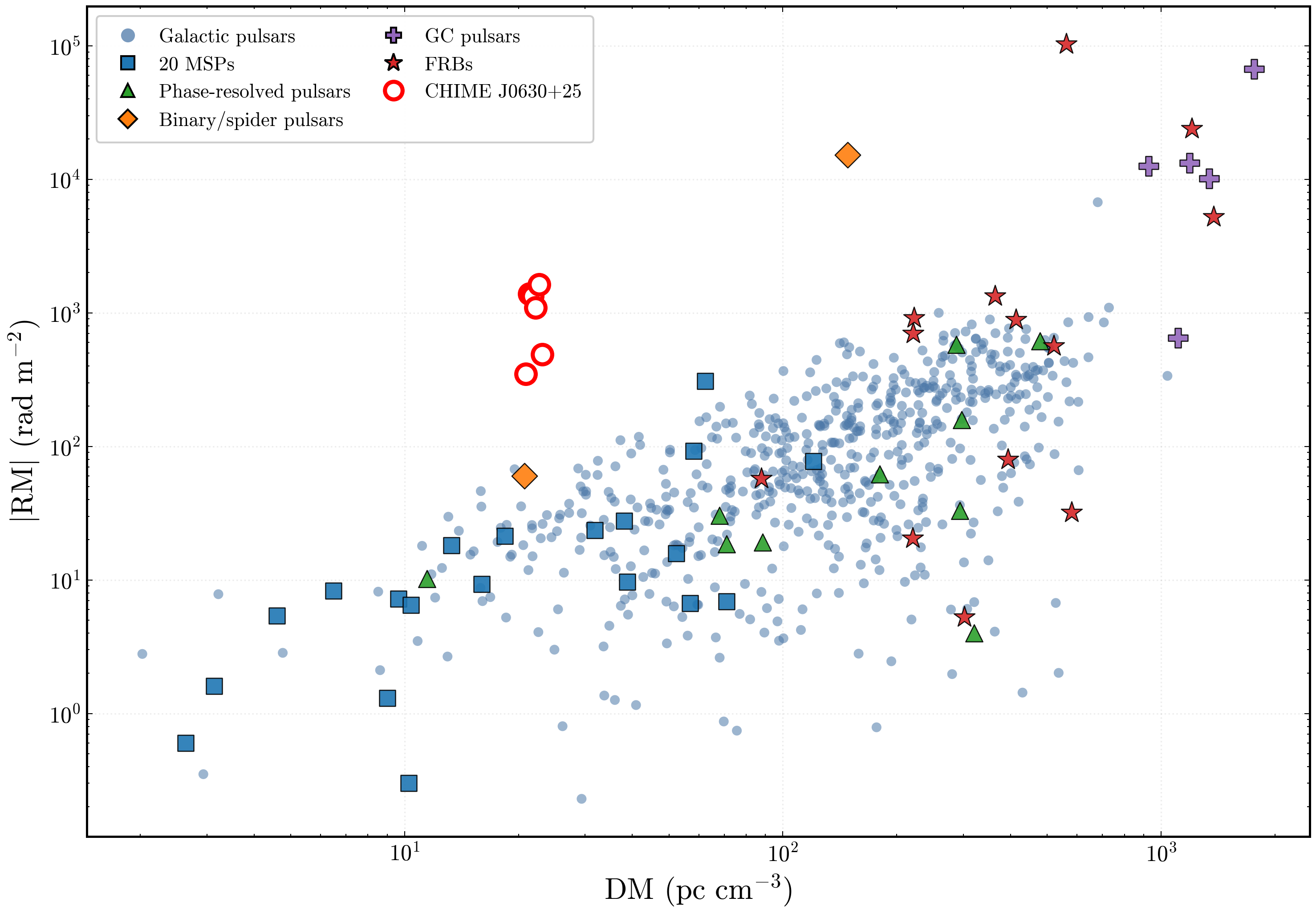}
    \caption{}
\end{subfigure}
\hfill
\begin{subfigure}{0.7\textwidth}
    \centering
    \includegraphics[width=\linewidth]{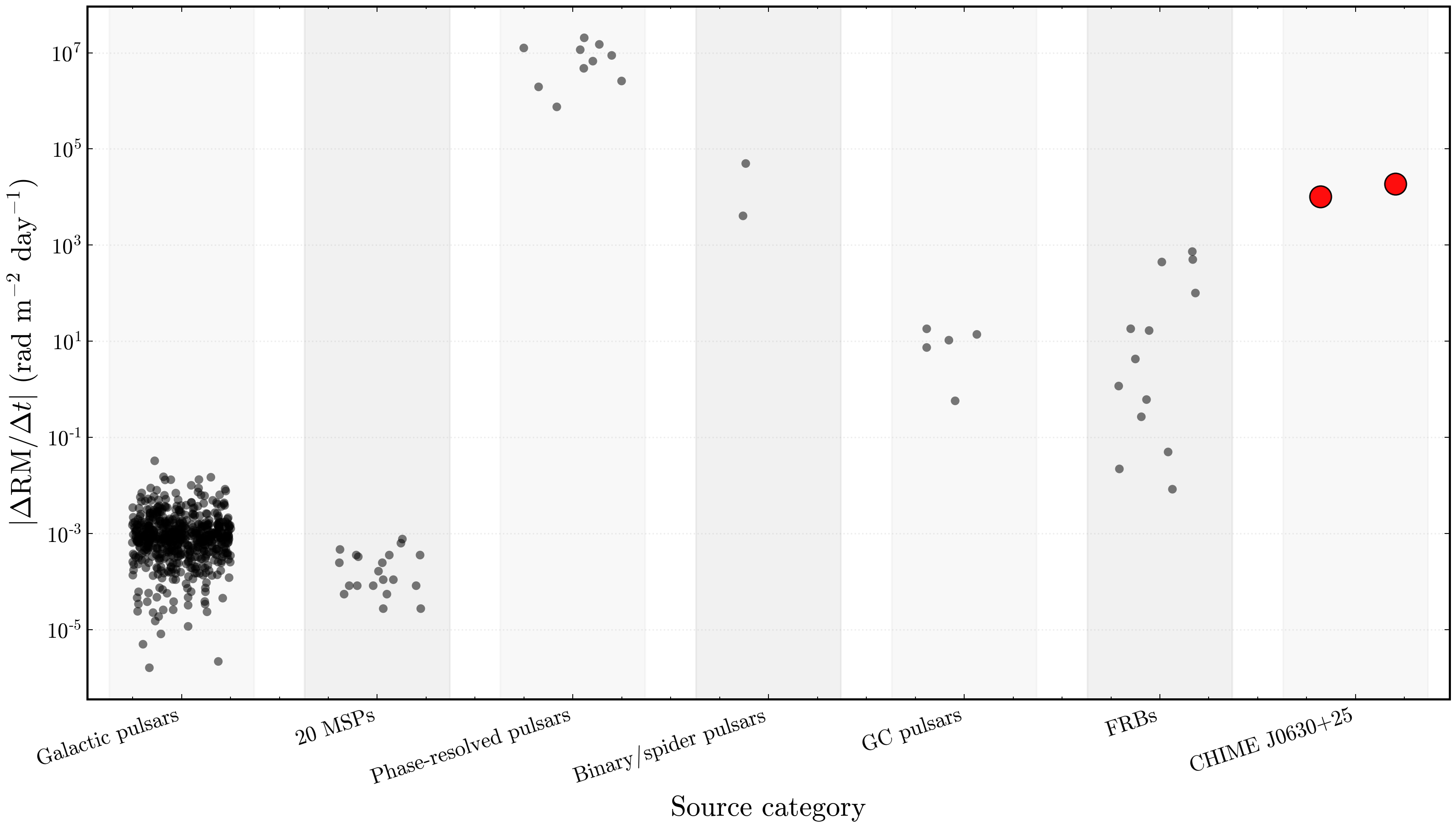}
    \caption{}
\end{subfigure}
\caption{
RM amplitude and variability of CHIME J0630$+$25 compared with other radio-emitting sources.
\textbf{(A)}: Absolute RM as a function of DM. Blue points show Galactic radio pulsars \cite{2024meerkat}, while coloured symbols denote 20 millisecond pulsars \cite{2011mspsr}, pulsars with phase-resolved RM variations \cite{2009phase}, binary/spider pulsars \cite{2002bpsr,2005psr1269,2023wangpsr}, Galactic-centre pulsars \cite{2023galacticcenter,2018gcp} and several repeating FRBs \cite{2021frb121102,2023chimefrb,2025Sfastrepeaterrm,2023frb,2022wangbebinary,202220200120E,202320220912}. Red open circles show all RM measurements of CHIME J0630$+$25. For comparison sources with multiple reported RM measurements, one representative point is plotted using its largest reported absolute RM value. The combination of very small ${\rm DM}\simeq22~{\rm pc~cm^{-3}}$ and large $|{\rm RM}|$ up to $1.6\times10^{3}~{\rm rad~m^{-2}}$ is unusual among radio sources and requires a magnetized Faraday screen local to the source. 
\textbf{(B)}: Largest reported RM variation rates for different source classes, using the same literature samples as in panel A. For phase-resolved pulsars, the values are apparent longitude-dependent RM gradients estimated as $\Delta{\rm RM}_{\rm pp}/P_{\rm spin}$ \cite{2009phase}, and are shown only for comparison. The two red points for CHIME J0630$+$25 correspond to the $\sim536~{\rm rad~m^{-2}}$ change within $\sim2500$ s and $\sim50~{\rm rad~m^{-2}}$ change between adjacent burst cycles.
}

\label{fig:rm_comparison}
\end{figure*}

To place CHIME J0630$+$25 in this broader context, we compare its RM amplitude and RM variation rate with those of several classes of radio-emitting sources in Fig.~\ref{fig:rm_comparison}. In the DM--$|{\rm RM}|$ plane, CHIME J0630$+$25 is clearly unusual: despite its very small ${\rm DM}\simeq22~{\rm pc~cm^{-3}}$, its $|{\rm RM}|$ reaches $\sim 1.6\times10^{3}~{\rm rad~m^{-2}}$, far above the ordinary Galactic pulsar population at comparable DM. In the RM-variation-rate comparison, CHIME J0630$+$25 also stands out. Its changes of $\sim50~{\rm rad~m^{-2}}$ between adjacent burst cycles and $\sim536~{\rm rad~m^{-2}}$ within $\sim2500$ s imply variation rates much larger than those of most ordinary pulsars and repeating FRBs. They are instead comparable to the most extreme RM changes seen in some binary or spider pulsar systems, where the radio signal propagates through dense and magnetized plasma local to the binary.
These binary pulsars provide another class of RM-variable sources, where the radio signal can propagate through magnetized plasma associated with the companion. 

In the black-widow system PSR J2051$-$0827, FAST observations revealed a regular RM decrease from $60$ to $-28.7~{\rm rad~m^{-2}}$ during eclipse egress, followed by recovery after the line of sight moved out of the eclipse medium \cite{2023wangpsr}. The Be X-ray binary PSR B1259$-$63 can exhibit RM variations of thousands of ${\rm rad~m^{-2}}$ over several days while DM remains nearly constant \cite{2005psr1269}.  These RM variations are attributed to the pulsar passing through the dense disk surrounding its Be-star companion. CHIME J0630$+$25 also shows a phase of depolarization, interpreted as arising from an extremely strong or disordered magnetic field or complex scattering. The rapid RM variations and depolarization therefore resemble binary pulsar systems in several aspects, although future observations are required to test whether an orbital period or companion is present.

One event on 2026-02-27 exhibited complex polarization behaviour as shown in Fig.~\ref{fig:observations}. At around $2644.25$ s, the Stokes $Q$ and $U$ dynamic spectra show a slanted pattern and a rapid jump, while Stokes $V$ reverses sign nearly simultaneously. The slanted $Q/U$ pattern could be caused by a time-dependent Faraday rotation, where lines of constant Faraday phase satisfy ${d{\rm RM}\over dt}=2\,{\rm RM}\,{1\over\nu}\,{d\nu\over dt}$. Using the measured slope of the stripes gives an effective RM drift rate of $3.3\times10^{3}~{\rm rad~m^{-2}~s^{-1}}$. Over several milliseconds, this corresponds to an RM change of tens of ${\rm rad~m^{-2}}$. 
However, we note that this feature can be also explained by a rapid intrinsic PA swing with constant RM followed by a polarization-mode transition resembling an orthogonal-polarization-mode-like transition. The simultaneous Stokes $V$ reversal supports this interpretation, because circular-polarization sign reversals associated with mode changes are commonly observed in radio pulsars \cite{1976opm,1978opm,1984opr,1998cp,2006cp2} and have also been reported in a LPT, GPM J1839-10 \cite{men2026detection}. We therefore favour an intrinsic polarization-angle change scenario for this event.

Therefore, CHIME J0630$+$25 is unusual in combining a low DM with large-amplitude and rapid RM variability. The observed RM changes cannot be explained by the ordinary interstellar medium. Instead, the Faraday rotation must be dominated by plasma local to the source, where the electron density, magnetic-field strength or line-of-sight geometry changes on timescales of minutes to hours. Its RM variability therefore traces a compact and rapidly evolving magneto-ionic environment, possibly associated with a binary system or with time-dependent magnetized plasma surrounding a compact object. Continued high-cadence full-Stokes monitoring and coordinated multi-telescope campaigns will be crucial for revealing the physical origin of this enigmatic long-period radio transient and for testing whether the RM evolution is stochastic, secular or periodic.

\clearpage

\clearpage
\newpage
\section*{Methods}

\section{FAST observations}

We observed CHIME J0630$+$25 with the Five-hundred-meter Aperture Spherical Telescope (FAST) in four epochs between 2025 August and 2026 March. The observations were carried out in tracking mode using the central beam of the 19-beam L-band receiver and the pulsar backend. Full-Stokes data were recorded with a time resolution of $98.304~\mu{\rm s}$ and 4096 frequency channels across the L-band observing band. A 10-K square-wave noise signal was injected at the beginning of each observation with a period of 0.2 s and was used for flux and polarization calibration. The observation times, durations and instrumental settings are listed in \ETab~\ref{tab:obs}.

All times quoted in this paper are measured relative to the start of each observation. The times of arrival of the detected pulses refer to the arrival time at the highest observing frequency, 1.5 GHz. We adopted ${\rm DM}=22~{\rm pc~cm^{-3}}$ throughout the analysis, consistent with the discovery report of CHIME J0630$+$25. We did not attempt a detailed DM optimization because the main results of this work rely on polarization and RM measurements rather than on absolute timing.

\section{Calibration}
The calibration procedure includes flux-density and polarization calibration. An example of the calibration results is shown in \EFig~\ref{fig:calres}.
\subsection{Flux-density calibration}

The raw data extracted from the FAST FITS files are recorded as power values in two orthogonal linear-polarization channels. We used the injected calibration signal to convert the raw power into antenna temperature. For the two polarization channels, the calibration factors were calculated as
\begin{align}
F_{{\rm cal},0} &= 
\frac{T_{{\rm cal},0}}
{P_{{\rm cal-on},0}-P_{{\rm cal-off},0}},\\
F_{{\rm cal},1} &= 
\frac{T_{{\rm cal},1}}
{P_{{\rm cal-on},1}-P_{{\rm cal-off},1}},
\end{align}
where $T_{\rm cal}$ is the standard noise-diode temperature, and $P_{\rm cal-on}$ and $P_{\rm cal-off}$ are the powers measured with and without the injected calibration signal. The calibrated antenna temperature was then calculated as
\begin{align}
T=\frac{P_0F_{{\rm cal},0}+P_1F_{{\rm cal},1}}{2}.
\end{align}

For each pulse, an off-pulse region adjacent to the burst was selected to estimate the local baseline. The pulse temperature was calculated as
\begin{align}
T_{\rm pulse}=T_{\rm ON}-T_{\rm OFF},
\end{align}
where $T_{\rm ON}$ and $T_{\rm OFF}$ denote the calibrated on-pulse and off-pulse temperatures. The corresponding flux density was obtained from
\begin{align}
S_{\rm pulse}=\frac{T_{\rm pulse}}{G},
\end{align}
where $G$ is the FAST gain at the corresponding zenith angle. Gain $G$ can be calculated either using calibration source, such as 3C 286, or through standard zenith-angle-dependent gain formula \cite{jiang2019commissioning,jiang2020fundamental}.

\subsection{Polarization calibration}

The recorded data contain the two linear-polarization intensities and the real and imaginary parts of the cross-correlation between the two polarization channels. These were converted into Stokes parameters $I$, $Q$, $U$ and $V$. We used the injected calibration signal to estimate the instrumental leakage, the relative phase delay between the two polarization channels, and the frequency-dependent normalization factor.

The leakage parameter was estimated as
\begin{align}
f_{\rm leak}=\frac{{\rm CAL}Q}{{\rm CAL}I},
\end{align}
where ${\rm CAL}$ prefixes represent the differences between the Stokes parameters measured with and without noise injection. The phase delay was calculated from
\begin{align}
\chi=\frac{1}{2}\arctan\frac{{\rm CAL}V}{{\rm CAL}U}.
\end{align}
The corrected Stokes parameters were then obtained by applying the leakage, phase-delay and normalization corrections to the observed Stokes parameters. The linear polarization intensity, total polarization intensity, linear polarization fraction, circular polarization fraction and polarization angle were calculated as
\begin{align}
L &= \sqrt{Q^2+U^2},\\
P &= \sqrt{L^2+V^2},\\
{\rm LP} &= \frac{L}{I},\\
{\rm CP} &= \frac{V}{I},\\
{\rm PA} &= \frac{1}{2}\arctan\frac{U}{-Q}.
\end{align}
The minus sign in the denominator of the PA definition was adopted to follow the IAU convention for the FAST polarization basis. Linear polarization was measured only when the polarized signal was detected with sufficient signal-to-noise ratio.

\section{RFI mitigation and pulse search}

Radio-frequency interference was removed using both automatic and manual procedures. For each observation, we calculated the off-pulse bandpass and fitted the spectral baseline using an asymmetrically reweighted penalized least-squares method \cite{baek2015baseline,zeng2021radio}, which was then subtracted from the bandpass. Channels with residual intensities above selected threshold were then flagged as RFI-contaminated. Remaining narrow-band or intermittent RFI features were inspected manually and removed before producing the final dynamic spectra.
 
Single pulses were searched using the PRESTO tools \texttt{prepsubband} and \texttt{single\_pulse\_search.py} \cite{ransom2011presto}.  Candidate events were grouped using a clustering procedure DBSCAN and filtered using peak scores \cite{pang2018novel}. For each detected burst, we measured the burst time range, peak flux density, polarization fractions and, when possible, the RM. The pulse properties are summarized in \ETab~\ref{tab:info}.

\section{Frequency-drift measurement}

Several bursts of CHIME J0630$+$25 show narrow-band structures whose central frequencies vary with time. We identified two types of apparent frequency-drifting behaviour: inter-pulse frequency drift, in which the central frequencies of multiple narrow-band single pulses vary within a burst cluster, and intra-pulse frequency drift, in which the emission drifts in frequency within a single pulse. For the inter-pulse drift, we estimated the drift rate by tracking the central frequency of each narrow-band pulse. For the intra-pulse drift, we selected two endpoints along the same component and estimated the apparent drift rate as
\begin{align}
\dot{\nu}\simeq\frac{\nu_{\rm B}-\nu_{\rm A}}{t_{\rm B}-t_{\rm A}},
\end{align}
where A and B denote the start and end points of the visually identified ridge. Positive and negative values of $\dot{\nu}$ correspond to upward and downward frequency drifts, respectively.

As shown in \EFig~\ref{fig:drift}, CHIME J0630$+$25 exhibits diverse drifting morphologies, including upward and downward drifts, as well as intra-pulse and inter-pulse drifts. We measured drift rates only for a few representative examples with clear time--frequency structures. The three intra-pulse features shown in \EFig~\ref{fig:drift}A have apparent drift rates of $1.0\times10^{5}$, $1.0\times10^{5}$ and $-5.1\times10^{4}~{\rm MHz~s^{-1}}$, respectively. The representative inter-pulse drift rate is of order $10^{4}~{\rm MHz~s^{-1}}$. These values are comparable in magnitude to the drifting rates reported for repeating FRBs \cite{2026frbdrift}. They may reflect intrinsic spectral evolution of the coherent emission or propagation through a structured plasma. Notably, no apparent DM variation could be identified among different RM-measurement epochs, mainly because of the low flux density and complex spectral morphology of the bursts. 

\section{RM calculation}
\subsection{RM synthesis}

We measured the rotation measure using RM synthesis from RM-tools \cite{purcell2020rm}. In RM synthesis, the complex linear polarization
\begin{align}
P(\lambda^2)=Q(\lambda^2)+iU(\lambda^2)
\end{align}
was derotated over a grid of trial RM values. For each trial RM, the derotated polarization vectors from different frequency channels were coherently summed. The best-fitting RM was taken to be the value that maximized the coherently derotated linear polarization intensity. This method avoids the $n\pi$ ambiguity that would arise from fitting PA directly as a function of $\lambda^2$. The RM uncertainty was estimated as the distance between the FDF peak and the linearly interpolated half-maximum point, divided by the signal-to-noise ratio of the FDF peak. RM-clean from RM-tools was applied where necessary to reduce sidelobes.

\subsection{$Q/U$ fitting}

For bursts with sufficiently high S/N, the frequency-dependent oscillations can be directly distinguished in dynamic spectra of Stokes parameters, as shown in \EFig~\ref{fig:IQUV}. For these pulses, we further fitted the frequency-dependent Stokes $Q$ and $U$ spectra directly to verify the RM value obtained through RM synthesis. The Stokes Q and U curves are fitted as:
\begin{align}
Q(\lambda^2) &= A\cos(2{\rm RM}\lambda^2)-B\sin(2{\rm RM}\lambda^2),\\
U(\lambda^2) &= A\sin(2{\rm RM}\lambda^2)+B\cos(2{\rm RM}\lambda^2),
\end{align}
where \(A\) and \(B\) are free parameters. Because the bursts show complex spectral structures, we did not normalize $Q$ and $U$ by the total linear polarization intensity before fitting and only attempted to fit the Faraday-rotation-induced oscillation. The agreement between the RM-synthesis results and the $Q/U$ fitting results was used as a consistency check for the high-S/N RM measurements. 

Additional RM-synthesis and \(Q/U\)-fitting examples are shown in \EFig~\ref{fig:rmcal_all}.

\subsection{Stokes parameters analysis}

For the burst detected near $t\simeq2644.25$ s on 2026 February 27, we examined the dynamic spectra of Stokes $Q$, $U$ and $V$ before and after Faraday derotation. The PA curve was calculated using the RM measured from RM synthesis. Some bursts, such as Fig.~\ref{fig:observations}D, exhibited complex spectra with frequency bands varying between single pulses. Thus, the PA values were calculated only within selected time--frequency windows, while the Stokes \(I\), \(L\) and \(V\) profiles were still formed using frequency range indicated by the yellow dashed lines. We used this burst to test two possible interpretations of the observed slanted $Q/U$ structures: time-dependent RM variation and intrinsic PA evolution.

If the slanted structures in \EFig~\ref{fig:qu} are interpreted as time-dependent Faraday rotation, the Faraday phase is
\begin{align}
\phi = 2{\rm RM}(t)\lambda^2.
\end{align}
For a line of constant phase, $d\phi/dt=0$, giving
\begin{align}
\frac{d{\rm RM}}{dt}
=
2{\rm RM}\frac{1}{\nu}\frac{d\nu}{dt}.
\end{align}
The measured stripe slope implies an effective RM drift rate of $\sim3.3\times10^{3}~{\rm rad~m^{-2}~s^{-1}}$. The millisecond-scale jumps are also present in Stokes $Q$ and $U$. Both features require an extremely compact emitting or propagation region. The nearly simultaneous reversal of Stokes $V$, however, favours an intrinsic PA swing and a transition between orthogonal polarization modes.

\section{QPO-like periodicity}

We searched for short-timescale quasi-periodic modulation in the Stokes $I$ light curves of the detected burst clusters. The frequency-averaged Stokes $I$ light curves were detrended to remove slow variations and then analyzed using autocorrelation functions, Fourier transforms and wavelet transforms.

As shown in \EFig~\ref{fig:qpo}, a common frequency of 3.644 Hz, corresponding to a period of 0.274 s, was identified in two burst cycles on 2026 February 6, 358.6--361.0 s and 5391--5392 s. Because this periodicity was detected in only two burst cycles, we treated it as a secondary property of CHIME J0630$+$25 rather than as the defining observational feature of the source. The modulation may arise from propagation through the same structured plasma responsible for the strong Faraday variability. Similar quasi-periodic or sub-pulse structures have been reported in radio-emitting neutron stars \cite{kramer2024quasi,chen2022discovery,tang2025quasi,tian2023subsecond}. This period of 0.274 s is also close to the characteristic sub-pulse timescale expected from the empirical relation $P_{\mu}\sim10^{-3}P$ reported for several pulsars, and may provide additional support for a neutron-star origin if similar periodicities are confirmed in future observations.

\section{Joint X-ray observations by HXMT}

Insight-HXMT is China's first X-ray astronomical satellite, launched on 2017 June 15. It carries three co-aligned collimated telescopes covering a broad X-ray energy range: the Low Energy X-ray telescope (LE; 1--15 keV), the Medium Energy X-ray telescope (ME; 5--30 keV) and the High Energy X-ray telescope (HE; 20--250 keV). The corresponding geometrical detector areas are approximately 384, 952 and 5,000 cm$^{2}$, respectively. The three instruments also provide fast timing capability, with time resolutions of 1 ms for LE, 280 $\mu$s for ME and 25 $\mu$s for HE, and energy resolutions of 2.5\% at 6 keV, 14\% at 20 keV and 19\% at 60 keV, respectively \cite{2020hxmt1}. 

CHIME J0630$+$25 was observed by \textit{Insight}-HXMT during three FAST observing sessions and one independent session. The \textit{Insight}-HXMT data were reduced with the Insight-HXMT Data Analysis Software (HXMTDAS) v2.06. Standard screening criteria were applied: the pointing offset angle was required to be less than $0.04^\circ$, the elevation angle greater than $10^\circ$, and the geomagnetic cutoff rigidity greater than 8 GV. Data taken within 300 s of passages through the South Atlantic Anomaly (SAA) were excluded. Light curves were extracted using the HXMTDAS tasks \texttt{helcgen}, \texttt{melcgen} and \texttt{lelcgen} for HE, ME and LE, respectively. The background spectra and light curves were estimated using the official background tools \texttt{HEBKGMAP}, \texttt{MEBKGMAP} and \texttt{LEBKGMAP}. The light curves were generated with a time bin of 0.0078125 s.

We calculated the background-subtracted average count rates for all four observations. No significant excess was detected in the instruments, with the average count rates being consistent with zero. This indicates that CHIME J0630$+$25 was not detected in X-rays during the \textit{Insight}-HXMT observations. To estimate the X-ray flux upper limits, we used the observation with the background-subtracted count rate and calculated the flux limits with the \texttt{flux} command in XSPEC. The resulting flux upper limits (3$\sigma$) are $3.3\times10^{-12}$, $6.0\times10^{-12}$ and $6.7\times10^{-12}~{\rm erg~cm^{-2}~s^{-1}}$ in the 2--10~keV (LE), 10--30~keV (ME) and 30--100~keV (HE) bands, respectively. Assuming isotropic emission and a distance of 170~pc, these correspond to isotropic luminosity upper limits of $1.1\times10^{31}$, $2.1\times10^{31}$ and $2.3\times10^{31}~{\rm erg~s^{-1}}$, respectively. The low X-ray luminosity disfavors a bright, persistently accreting neutron-star X-ray binary interpretation.



\bibliographystyle{naturemag}

\bibliography{na2}

@ARTICLE{rea2026long,
       author = {{Rea}, Nanda and {Hurley-Walker}, Natasha and {Caleb}, Manisha},
        title = "{Long period transients (LPTs): A comprehensive review}",
      journal = {Journal of High Energy Astrophysics},
         year = 2026,
        month = apr,
       volume = {52},
          eid = {100566},
        pages = {100566},
          doi = {10.1016/j.jheap.2026.100566},
archivePrefix = {arXiv},
       eprint = {2601.10393},
 primaryClass = {astro-ph.HE},
       adsurl = {https://ui.adsabs.harvard.edu/abs/2026JHEAp..5200566R}
}

@ARTICLE{zelati2024ultra,
       author = {{Coti Zelati}, Francesco and {Borghese}, Alice},
        title = "{Ultra-long period compact sources: a glimpse into observational breakthroughs and theoretical challenges}",
      journal = {arXiv e-prints},
         year = 2024,
        month = dec,
          eid = {arXiv:2412.12763},
        pages = {arXiv:2412.12763},
          doi = {10.48550/arXiv.2412.12763},
archivePrefix = {arXiv},
       eprint = {2412.12763},
 primaryClass = {astro-ph.HE},
       adsurl = {https://ui.adsabs.harvard.edu/abs/2024arXiv241212763C}
}

@ARTICLE{lee2025emission,
       author = {{Lee}, Y.~W.~J. and {Caleb}, M. and {Murphy}, Tara and {Lenc}, E. and {Kaplan}, D.~L. and {Ferrario}, L. and {Wadiasingh}, Z. and {Anumarlapudi}, A. and {Hurley-Walker}, N. and {Karambelkar}, V. and {Ocker}, S.~K. and {McSweeney}, S. and {Qiu}, H. and {Rajwade}, K.~M. and {Zic}, A. and {Bannister}, K.~W. and {Bhat}, N.~D.~R. and {Deller}, A. and {Dobie}, D. and {Driessen}, L.~N. and {Gendreau}, K. and {Glowacki}, M. and {Gupta}, V. and {Jahns-Schindler}, J.~N. and {Jaini}, A. and {James}, C.~W. and {Kasliwal}, M.~M. and {Lower}, M.~E. and {Shannon}, R.~M. and {Uttarkar}, P.~A. and {Wang}, Y. and {Wang}, Z.},
        title = "{The emission of interpulses by a 6.45-h-period coherent radio transient}",
      journal = {Nature Astronomy},
         year = 2025,
        month = mar,
       volume = {9},
        pages = {393-405},
          doi = {10.1038/s41550-024-02452-z},
archivePrefix = {arXiv},
       eprint = {2501.09133},
 primaryClass = {astro-ph.HE},
       adsurl = {https://ui.adsabs.harvard.edu/abs/2025NatAs...9..393L}
}

@ARTICLE{hurley2023long,
       author = {{Hurley-Walker}, N. and {Rea}, N. and {McSweeney}, S.~J. and {Meyers}, B.~W. and {Lenc}, E. and {Heywood}, I. and {Hyman}, S.~D. and {Men}, Y.~P. and {Clarke}, T.~E. and {Coti Zelati}, F. and {Price}, D.~C. and {Horv{\'a}th}, C. and {Galvin}, T.~J. and {Anderson}, G.~E. and {Bahramian}, A. and {Barr}, E.~D. and {Bhat}, N.~D.~R. and {Caleb}, M. and {Dall'Ora}, M. and {de Martino}, D. and {Giacintucci}, S. and {Morgan}, J.~S. and {Rajwade}, K.~M. and {Stappers}, B. and {Williams}, A.},
        title = "{A long-period radio transient active for three decades}",
      journal = {\nat},
         year = 2023,
        month = jul,
       volume = {619},
       number = {7970},
        pages = {487-490},
          doi = {10.1038/s41586-023-06202-5},
archivePrefix = {arXiv},
       eprint = {2503.08036},
 primaryClass = {astro-ph.HE},
       adsurl = {https://ui.adsabs.harvard.edu/abs/2023Natur.619..487H}
}

@ARTICLE{men2025highly,
       author = {{Men}, Yunpeng and {McSweeney}, Sam and {Hurley-Walker}, Natasha and {Barr}, Ewan and {Stappers}, Ben},
        title = "{A highly magnetized long-period radio transient exhibiting unusual emission features}",
      journal = {Science Advances},
         year = 2025,
        month = jan,
       volume = {11},
       number = {3},
          eid = {eadp6351},
        pages = {eadp6351},
          doi = {10.1126/sciadv.adp6351},
archivePrefix = {arXiv},
       eprint = {2501.10528},
 primaryClass = {astro-ph.HE},
       adsurl = {https://ui.adsabs.harvard.edu/abs/2025SciA...11P6351M}
}

@ARTICLE{caleb2024emission,
       author = {{Caleb}, M. and {Lenc}, E. and {Kaplan}, D.~L. and {Murphy}, T. and {Men}, Y.~P. and {Shannon}, R.~M. and {Ferrario}, L. and {Rajwade}, K.~M. and {Clarke}, T.~E. and {Giacintucci}, S. and {Hurley-Walker}, N. and {Hyman}, S.~D. and {Lower}, M.~E. and {McSweeney}, Sam and {Ravi}, V. and {Barr}, E.~D. and {Buchner}, S. and {Flynn}, C.~M.~L. and {Hessels}, J.~W.~T. and {Kramer}, M. and {Pritchard}, J. and {Stappers}, B.~W.},
        title = "{An emission-state-switching radio transient with a 54-minute period}",
      journal = {Nature Astronomy},
         year = 2024,
        month = sep,
       volume = {8},
        pages = {1159-1168},
          doi = {10.1038/s41550-024-02277-w},
archivePrefix = {arXiv},
       eprint = {2407.12266},
 primaryClass = {astro-ph.HE},
       adsurl = {https://ui.adsabs.harvard.edu/abs/2024NatAs...8.1159C}
}

@ARTICLE{wang2025detection,
       author = {{Wang}, Ziteng and {Rea}, Nanda and {Bao}, Tong and {Kaplan}, David L. and {Lenc}, Emil and {Wadiasingh}, Zorawar and {Hare}, Jeremy and {Zic}, Andrew and {Anumarlapudi}, Akash and {Bera}, Apurba and {Beniamini}, Paz and {Cooper}, A.~J. and {Clarke}, Tracy E. and {Deller}, Adam T. and {Dawson}, J.~R. and {Glowacki}, Marcin and {Hurley-Walker}, Natasha and {McSweeney}, S.~J. and {Polisensky}, Emil J. and {Peters}, Wendy M. and {Younes}, George and {Bannister}, Keith W. and {Caleb}, Manisha and {Dage}, Kristen C. and {James}, Clancy W. and {Kasliwal}, Mansi M. and {Karambelkar}, Viraj and {Lower}, Marcus E. and {Mori}, Kaya and {Ocker}, Stella Koch and {P{\'e}rez-Torres}, Miguel and {Qiu}, Hao and {Rose}, Kovi and {Shannon}, Ryan M. and {Taub}, Rhianna and {Wang}, Fayin and {Wang}, Yuanming and {Zhao}, Zhenyin and {Bhat}, N.~D. Ramesh and {Dobie}, Dougal and {Driessen}, Laura N. and {Murphy}, Tara and {Jaini}, Akhil and {Deng}, Xinping and {Jahns-Schindler}, Joscha N. and {Lee}, Y.~W. Joshua and {Pritchard}, Joshua and {Tuthill}, John and {Thyagarajan}, Nithyanandan},
        title = "{Detection of X-ray emission from a bright long-period radio transient}",
      journal = {\nat},
         year = 2025,
        month = jun,
       volume = {642},
       number = {8068},
        pages = {583-586},
          doi = {10.1038/s41586-025-09077-w},
archivePrefix = {arXiv},
       eprint = {2411.16606},
 primaryClass = {astro-ph.HE},
       adsurl = {https://ui.adsabs.harvard.edu/abs/2025Natur.642..583W}
}

@ARTICLE{dong2025chime,
       author = {{Dong}, Fengqiu Adam and {Clarke}, Tracy E. and {Curtin}, Alice and {Kumar}, Ajay and {Mckinven}, Ryan and {Shin}, Kaitlyn and {Stairs}, Ingrid and {Brar}, Charanjot and {Burdge}, Kevin and {Chatterjee}, Shami and {Cook}, Amanda M. and {Fonseca}, Emmanuel and {Gaensler}, B.~M. and {Hessels}, Jason W. and {Kaspi}, Victoria M. and {Lazda}, Mattias and {Main}, Robert and {Masui}, Kiyoshi W. and {McKee}, James W. and {Meyers}, Bradley W. and {Pearlman}, Aaron B. and {Ransom}, Scott M. and {Scholz}, Paul and {Smith}, Kendrick M. and {Tan}, Chia Min},
        title = "{CHIME/Fast Radio Burst/Pulsar Discovery of a Nearby Long-period Radio Transient with a Timing Glitch}",
      journal = {\apjl},
         year = 2025,
        month = sep,
       volume = {990},
       number = {2},
          eid = {L49},
        pages = {L49},
          doi = {10.3847/2041-8213/adfa8e},
archivePrefix = {arXiv},
       eprint = {2407.07480},
 primaryClass = {astro-ph.HE},
       adsurl = {https://ui.adsabs.harvard.edu/abs/2025ApJ...990L..49D}
}

@ARTICLE{jiang2019commissioning,
       author = {{Jiang}, Peng and {Yue}, YouLing and {Gan}, HengQian and {Yao}, Rui and {Li}, Hui and {Pan}, GaoFeng and {Sun}, JingHai and {Yu}, DongJun and {Liu}, HongFei and {Tang}, NingYu and {Qian}, Lei and {Lu}, JiGuang and {Yan}, Jun and {Peng}, Bo and {Zhang}, ShuXin and {Wang}, QiMing and {Li}, Qi and {Li}, Di and {FAST Collaboration}},
        title = "{Commissioning progress of the FAST}",
      journal = {Science China Physics, Mechanics, and Astronomy},
         year = 2019,
        month = may,
       volume = {62},
       number = {5},
          eid = {959502},
        pages = {959502},
          doi = {10.1007/s11433-018-9376-1},
archivePrefix = {arXiv},
       eprint = {1903.06324},
 primaryClass = {astro-ph.IM},
       adsurl = {https://ui.adsabs.harvard.edu/abs/2019SCPMA..6259502J}
}

@ARTICLE{jiang2020fundamental,
       author = {{Jiang}, Peng and {Tang}, Ning-Yu and {Hou}, Li-Gang and {Liu}, Meng-Ting and {Kr{\v{c}}o}, Marko and {Qian}, Lei and {Sun}, Jing-Hai and {Ching}, Tao-Chung and {Liu}, Bin and {Duan}, Yan and {Yue}, You-Ling and {Gan}, Heng-Qian and {Yao}, Rui and {Li}, Hui and {Pan}, Gao-Feng and {Yu}, Dong-Jun and {Liu}, Hong-Fei and {Li}, Di and {Peng}, Bo and {Yan}, Jun and {FAST Collaboration}},
        title = "{The fundamental performance of FAST with 19-beam receiver at L band}",
      journal = {Research in Astronomy and Astrophysics},
         year = 2020,
        month = may,
       volume = {20},
       number = {5},
          eid = {064},
        pages = {064},
          doi = {10.1088/1674-4527/20/5/64},
archivePrefix = {arXiv},
       eprint = {2002.01786},
 primaryClass = {astro-ph.IM},
       adsurl = {https://ui.adsabs.harvard.edu/abs/2020RAA....20...64J}
}

@ARTICLE{pang2018novel,
       author = {{Pang}, Di and {Goseva-Popstojanova}, Katerina and {Devine}, Thomas and {McLaughlin}, Maura},
        title = "{A novel single-pulse search approach to detection of dispersed radio pulses using clustering and supervised machine learning}",
      journal = {\mnras},
         year = 2018,
        month = nov,
       volume = {480},
       number = {3},
        pages = {3302-3323},
          doi = {10.1093/mnras/sty1992},
archivePrefix = {arXiv},
       eprint = {1807.07164},
 primaryClass = {astro-ph.IM},
       adsurl = {https://ui.adsabs.harvard.edu/abs/2018MNRAS.480.3302P}
}

@ARTICLE{zeng2021radio,
       author = {{Zeng}, Qingguo and {Chen}, Xue and {Li}, Xiangru and {Han}, J.~L. and {Wang}, Chen and {Zhou}, D.~J. and {Wang}, Tao},
        title = "{Radio frequency interference mitigation based on the asymmetrically reweighted penalized least squares and SumThreshold method}",
      journal = {\mnras},
         year = 2021,
        month = jan,
       volume = {500},
       number = {3},
        pages = {2969-2978},
          doi = {10.1093/mnras/staa2551},
archivePrefix = {arXiv},
       eprint = {2008.11949},
 primaryClass = {astro-ph.IM},
       adsurl = {https://ui.adsabs.harvard.edu/abs/2021MNRAS.500.2969Z}
}

@ARTICLE{baek2015baseline,
       author = {{Baek}, Sung-June and {Park}, Aaron and {Ahn}, Young-Jin and {Choo}, Jaebum},
        title = "{Baseline correction using asymmetrically reweighted penalized least squares smoothing}",
      journal = {The Analyst},
         year = 2015,
        month = jan,
       volume = {140},
       number = {1},
        pages = {250-257},
          doi = {10.1039/C4AN01061B},
       adsurl = {https://ui.adsabs.harvard.edu/abs/2015Ana...140..250B}
}

@ARTICLE{kramer2024quasi,
       author = {{Kramer}, Michael and {Liu}, Kuo and {Desvignes}, Gregory and {Karuppusamy}, Ramesh and {Stappers}, Ben W.},
        title = "{Quasi-periodic sub-pulse structure as a unifying feature for radio-emitting neutron stars}",
      journal = {\nastro},
         year = 2024,
        month = feb,
       volume = {8},
        pages = {230-240},
          doi = {10.1038/s41550-023-02125-3},
archivePrefix = {arXiv},
       eprint = {2311.13762},
 primaryClass = {astro-ph.HE},
       adsurl = {https://ui.adsabs.harvard.edu/abs/2024NatAs...8..230K}
}

@ARTICLE{chen2022discovery,
       author = {{Chen}, J.~L. and {Wen}, Z.~G. and {Yuan}, J.~P. and {Wang}, N. and {Li}, D. and {Wang}, H.~G. and {Yan}, W.~M. and {Yuen}, R. and {Wang}, P. and {Wang}, Z. and {Zhu}, W.~W. and {Niu}, J.~R. and {Miao}, C.~C. and {Xue}, M.~Y. and {Gong}, B.~P.},
        title = "{The Discovery of a Rotating Radio Transient J1918-0449 with Intriguing Emission Properties with the Five-hundred-meter Aperture Spherical Radio Telescope}",
      journal = {\apj},
         year = 2022,
        month = jul,
       volume = {934},
       number = {1},
          eid = {24},
        pages = {24},
          doi = {10.3847/1538-4357/ac75d1},
archivePrefix = {arXiv},
       eprint = {2206.03091},
 primaryClass = {astro-ph.HE},
       adsurl = {https://ui.adsabs.harvard.edu/abs/2022ApJ...934...24C}
}

@ARTICLE{tang2025quasi,
       author = {{Tang}, Zhenfan and {Zhang}, Songbo and {Wang}, Jieshuang and {Yang}, Xuan and {Wu}, Xuefeng},
        title = "{Quasi-periodic sub-structure of RRAT J1913+1330}",
      journal = {\mnras},
         year = 2025,
        month = may,
       volume = {539},
       number = {2},
        pages = {1352-1358},
          doi = {10.1093/mnras/staf570},
archivePrefix = {arXiv},
       eprint = {2504.04985},
 primaryClass = {astro-ph.HE},
       adsurl = {https://ui.adsabs.harvard.edu/abs/2025MNRAS.539.1352T}
}

@ARTICLE{tian2023subsecond,
       author = {{Tian}, Pengfu and {Zhang}, Ping and {Wang}, Wei and {Wang}, Pei and {Sun}, Xiaohui and {Liu}, Jifeng and {Zhang}, Bing and {Dai}, Zigao and {Yuan}, Feng and {Zhang}, Shuangnan and {Liu}, Qingzhong and {Jiang}, Peng and {Wu}, Xuefeng and {Zheng}, Zheng and {Chen}, Jiashi and {Li}, Di and {Zhu}, Zonghong and {Pan}, Zhichen and {Gan}, Hengqian and {Chen}, Xiao and {Sai}, Na},
        title = "{Subsecond periodic radio oscillations in a microquasar}",
      journal = {\nat},
         year = 2023,
        month = sep,
       volume = {621},
       number = {7978},
        pages = {271-275},
          doi = {10.1038/s41586-023-06336-6},
archivePrefix = {arXiv},
       eprint = {2307.14015},
 primaryClass = {astro-ph.HE},
       adsurl = {https://ui.adsabs.harvard.edu/abs/2023Natur.621..271T}
}

@software{ransom2011presto,
       author = {{Ransom}, Scott},
        title = "{PRESTO: PulsaR Exploration and Search TOolkit}",
 howpublished = {Astrophysics Source Code Library, record ascl:1107.017},
         year = 2011,
        month = jul,
          eid = {ascl:1107.017},
archivePrefix = {ascl},
       eprint = {1107.017},
       adsurl = {https://ui.adsabs.harvard.edu/abs/2011ascl.soft07017R}
}

@ARTICLE{hutschenreuter2022galactic,
       author = {{Hutschenreuter}, S. and {Anderson}, C.~S. and {Betti}, S. and {Bower}, G.~C. and {Brown}, J.-A. and {Br{\"u}ggen}, M. and {Carretti}, E. and {Clarke}, T. and {Clegg}, A. and {Costa}, A. and {Croft}, S. and {Van Eck}, C. and {Gaensler}, B.~M. and {de Gasperin}, F. and {Haverkorn}, M. and {Heald}, G. and {Hull}, C.~L.~H. and {Inoue}, M. and {Johnston-Hollitt}, M. and {Kaczmarek}, J. and {Law}, C. and {Ma}, Y.~K. and {MacMahon}, D. and {Mao}, S.~A. and {Riseley}, C. and {Roy}, S. and {Shanahan}, R. and {Shimwell}, T. and {Stil}, J. and {Sobey}, C. and {O'Sullivan}, S.~P. and {Tasse}, C. and {Vacca}, V. and {Vernstrom}, T. and {Williams}, P.~K.~G. and {Wright}, M. and {En{\ss}lin}, T.~A.},
        title = "{The Galactic Faraday rotation sky 2020}",
      journal = {\aap},
         year = 2022,
        month = jan,
       volume = {657},
          eid = {A43},
        pages = {A43},
          doi = {10.1051/0004-6361/202140486},
archivePrefix = {arXiv},
       eprint = {2102.01709},
 primaryClass = {astro-ph.GA},
       adsurl = {https://ui.adsabs.harvard.edu/abs/2022A&A...657A..43H}
}

@ARTICLE{1994han,
       author = {{Han}, J.~L. and {Qiao}, G.~J.},
        title = "{The magnetic field in the disk of our Galaxy}",
      journal = {\aap},
         year = 1994,
        month = aug,
       volume = {288},
        pages = {759-772},
       adsurl = {https://ui.adsabs.harvard.edu/abs/1994A&A...288..759H}
}

@ARTICLE{2006han,
       author = {{Han}, J.~L. and {Manchester}, R.~N. and {Lyne}, A.~G. and {Qiao}, G.~J. and {van Straten}, W.},
        title = "{Pulsar Rotation Measures and the Large-Scale Structure of the Galactic Magnetic Field}",
      journal = {\apj},
         year = 2006,
        month = may,
       volume = {642},
       number = {2},
        pages = {868-881},
          doi = {10.1086/501444},
archivePrefix = {arXiv},
       eprint = {astro-ph/0601357},
 primaryClass = {astro-ph},
       adsurl = {https://ui.adsabs.harvard.edu/abs/2006ApJ...642..868H}
}

@ARTICLE{2022wangbebinary,
       author = {{Wang}, F.~Y. and {Zhang}, G.~Q. and {Dai}, Z.~G. and {Cheng}, K.~S.},
        title = "{Repeating fast radio burst 20201124A originates from a magnetar/Be star binary}",
      journal = {Nature Communications},
         year = 2022,
        month = sep,
       volume = {13},
          eid = {4382},
        pages = {4382},
          doi = {10.1038/s41467-022-31923-y},
archivePrefix = {arXiv},
       eprint = {2204.08124},
 primaryClass = {astro-ph.HE},
       adsurl = {https://ui.adsabs.harvard.edu/abs/2022NatCo..13.4382W}
}

@ARTICLE{2023galacticcenter,
       author = {{Abbate}, F. and {Noutsos}, A. and {Desvignes}, G. and {Wharton}, R.~S. and {Torne}, P. and {Kramer}, M. and {Eatough}, R.~P. and {Karuppusamy}, R. and {Liu}, K. and {Shao}, L. and {Wongphechauxsorn}, J.},
        title = "{Rotation measure variations in Galactic Centre pulsars}",
      journal = {\mnras},
         year = 2023,
        month = sep,
       volume = {524},
       number = {2},
        pages = {2966-2977},
          doi = {10.1093/mnras/stad2047},
archivePrefix = {arXiv},
       eprint = {2307.03230},
 primaryClass = {astro-ph.HE},
       adsurl = {https://ui.adsabs.harvard.edu/abs/2023MNRAS.524.2966A}
}

@ARTICLE{2007XTE,
       author = {{Camilo}, F. and {Reynolds}, J. and {Johnston}, S. and {Halpern}, J.~P. and {Ransom}, S.~M. and {van Straten}, W.},
        title = "{Polarized Radio Emission from the Magnetar XTE J1810-197}",
      journal = {\apjl},
         year = 2007,
        month = apr,
       volume = {659},
       number = {1},
        pages = {L37-L40},
          doi = {10.1086/516630},
archivePrefix = {arXiv},
       eprint = {astro-ph/0702616},
 primaryClass = {astro-ph},
       adsurl = {https://ui.adsabs.harvard.edu/abs/2007ApJ...659L..37C}
}

@ARTICLE{2009phase,
       author = {{Noutsos}, A. and {Karastergiou}, A. and {Kramer}, M. and {Johnston}, S. and {Stappers}, B.~W.},
        title = "{Phase-resolved Faraday rotation in pulsars}",
      journal = {\mnras},
         year = 2009,
        month = jul,
       volume = {396},
       number = {3},
        pages = {1559-1572},
          doi = {10.1111/j.1365-2966.2009.14806.x},
archivePrefix = {arXiv},
       eprint = {0903.5511},
 primaryClass = {astro-ph.GA},
       adsurl = {https://ui.adsabs.harvard.edu/abs/2009MNRAS.396.1559N}
}

@ARTICLE{2019XTE,
       author = {{Dai}, Shi and {Lower}, Marcus E. and {Bailes}, Matthew and {Camilo}, Fernando and {Halpern}, Jules P. and {Johnston}, Simon and {Kerr}, Matthew and {Reynolds}, John and {Sarkissian}, John and {Scholz}, Paul},
        title = "{Wideband Polarized Radio Emission from the Newly Revived Magnetar XTE J1810-197}",
      journal = {\apjl},
         year = 2019,
        month = apr,
       volume = {874},
       number = {2},
          eid = {L14},
        pages = {L14},
          doi = {10.3847/2041-8213/ab0e7a},
archivePrefix = {arXiv},
       eprint = {1902.04689},
 primaryClass = {astro-ph.HE},
       adsurl = {https://ui.adsabs.harvard.edu/abs/2019ApJ...874L..14D}
}

@ARTICLE{2006XTE,
       author = {{Camilo}, Fernando and {Ransom}, Scott M. and {Halpern}, Jules P. and {Reynolds}, John and {Helfand}, David J. and {Zimmerman}, Neil and {Sarkissian}, John},
        title = "{Transient pulsed radio emission from a magnetar}",
      journal = {\nat},
         year = 2006,
        month = aug,
       volume = {442},
       number = {7105},
        pages = {892-895},
          doi = {10.1038/nature04986},
archivePrefix = {arXiv},
       eprint = {astro-ph/0605429},
 primaryClass = {astro-ph},
       adsurl = {https://ui.adsabs.harvard.edu/abs/2006Natur.442..892C}
}

@ARTICLE{2021swift,
       author = {{Lower}, M.~E. and {Johnston}, S. and {Shannon}, R.~M. and {Bailes}, M. and {Camilo}, F.},
        title = "{The dynamic magnetosphere of Swift J1818.0-1607}",
      journal = {\mnras},
         year = 2021,
        month = mar,
       volume = {502},
       number = {1},
        pages = {127-139},
          doi = {10.1093/mnras/staa3789},
archivePrefix = {arXiv},
       eprint = {2011.12463},
 primaryClass = {astro-ph.HE},
       adsurl = {https://ui.adsabs.harvard.edu/abs/2021MNRAS.502..127L}
}

@ARTICLE{2012psr,
       author = {{Levin}, L. and {Bailes}, M. and {Bates}, S.~D. and {Bhat}, N.~D.~R. and {Burgay}, M. and {Burke-Spolaor}, S. and {D'Amico}, N. and {Johnston}, S. and {Keith}, M.~J. and {Kramer}, M. and {Milia}, S. and {Possenti}, A. and {Stappers}, B. and {van Straten}, W.},
        title = "{Radio emission evolution, polarimetry and multifrequency single pulse analysis of the radio magnetar PSR J1622-4950}",
      journal = {\mnras},
         year = 2012,
        month = may,
       volume = {422},
       number = {3},
        pages = {2489-2500},
          doi = {10.1111/j.1365-2966.2012.20807.x},
archivePrefix = {arXiv},
       eprint = {1204.2045},
 primaryClass = {astro-ph.HE},
       adsurl = {https://ui.adsabs.harvard.edu/abs/2012MNRAS.422.2489L}
}

@ARTICLE{2018psr,
       author = {{Camilo}, F. and {Scholz}, P. and {Serylak}, M. and {Buchner}, S. and {Merryfield}, M. and {Kaspi}, V.~M. and {Archibald}, R.~F. and {Bailes}, M. and {Jameson}, A. and {van Straten}, W. and {Sarkissian}, J. and {Reynolds}, J.~E. and {Johnston}, S. and {Hobbs}, G. and {Abbott}, T.~D. and {Adam}, R.~M. and {Adams}, G.~B. and {Alberts}, T. and {Andreas}, R. and {Asad}, K.~M.~B. and {Baker}, D.~E. and {Baloyi}, T. and {Bauermeister}, E.~F. and {Baxana}, T. and {Bennett}, T.~G.~H. and {Bernardi}, G. and {Booisen}, D. and {Booth}, R.~S. and {Botha}, D.~H. and {Boyana}, L. and {Brederode}, L.~R.~S. and {Burger}, J.~P. and {Cheetham}, T. and {Conradie}, J. and {Conradie}, J.~P. and {Davidson}, D.~B. and {De Bruin}, G. and {de Swardt}, B. and {de Villiers}, C. and {de Villiers}, D.~I.~L. and {de Villiers}, M.~S. and {de Villiers}, W. and {De Waal}, C. and {Dikgale}, M.~A. and {du Toit}, G. and {du Toit}, L.~J. and {Esterhuyse}, S.~W.~P. and {Fanaroff}, B. and {Fataar}, S. and {Foley}, A.~R. and {Foster}, G. and {Fourie}, D. and {Gamatham}, R. and {Gatsi}, T. and {Geschke}, R. and {Goedhart}, S. and {Grobler}, T.~L. and {Gumede}, S.~C. and {Hlakola}, M.~J. and {Hokwana}, A. and {Hoorn}, D.~H. and {Horn}, D. and {Horrell}, J. and {Hugo}, B. and {Isaacson}, A. and {Jacobs}, O. and {Jansen van Rensburg}, J.~P. and {Jonas}, J.~L. and {Jordaan}, B. and {Joubert}, A. and {Joubert}, F. and {J{\'o}zsa}, G.~I.~G. and {Julie}, R. and {Julius}, C.~C. and {Kapp}, F. and {Karastergiou}, A. and {Karels}, F. and {Kariseb}, M. and {Karuppusamy}, R. and {Kasper}, V. and {Knox-Davies}, E.~C. and {Koch}, D. and {Kotz{\'e}}, P.~P.~A. and {Krebs}, A. and {Kriek}, N. and {Kriel}, H. and {Kusel}, T. and {Lamoor}, S. and {Lehmensiek}, R. and {Liebenberg}, D. and {Liebenberg}, I. and {Lord}, R.~T. and {Lunsky}, B. and {Mabombo}, N. and {Macdonald}, T. and {Macfarlane}, P. and {Madisa}, K. and {Mafhungo}, L. and {Magnus}, L.~G. and {Magozore}, C. and {Mahgoub}, O. and {Main}, J.~P.~L. and {Makhathini}, S. and {Malan}, J.~A. and {Malgas}, P. and {Manley}, J.~R. and {Manzini}, M. and {Marais}, L. and {Marais}, N. and {Marais}, S.~J. and {Maree}, M. and {Martens}, A. and {Matshawule}, S.~D. and {Matthysen}, N. and {Mauch}, T. and {McNally}, L.~D. and {Merry}, B. and {Millenaar}, R.~P. and {Mjikelo}, C. and {Mkhabela}, N. and {Mnyandu}, N. and {Moeng}, I.~T. and {Mokone}, O.~J. and {Monama}, T.~E. and {Montshiwa}, K. and {Moss}, V. and {Mphego}, M. and {New}, W. and {Ngcebetsha}, B. and {Ngoasheng}, K. and {Niehaus}, H. and {Ntuli}, P. and {Nzama}, A. and {Obies}, F. and {Obrocka}, M. and {Ockards}, M.~T. and {Olyn}, C. and {Oozeer}, N. and {Otto}, A.~J. and {Padayachee}, Y. and {Passmoor}, S. and {Patel}, A.~A. and {Paula}, S. and {Peens-Hough}, A. and {Pholoholo}, B. and {Prozesky}, P. and {Rakoma}, S. and {Ramaila}, A.~J.~T. and {Rammala}, I. and {Ramudzuli}, Z.~R. and {Rasivhaga}, M. and {Ratcliffe}, S. and {Reader}, H.~C. and {Renil}, R. and {Richter}, L. and {Robyntjies}, A. and {Rosekrans}, D. and {Rust}, A. and {Salie}, S. and {Sambu}, N. and {Schollar}, C.~T.~G. and {Schwardt}, L. and {Seranyane}, S. and {Sethosa}, G. and {Sharpe}, C. and {Siebrits}, R. and {Sirothia}, S.~K. and {Slabber}, M.~J. and {Smirnov}, O. and {Smith}, S. and {Sofeya}, L. and {Songqumase}, N. and {Spann}, R. and {Stappers}, B. and {Steyn}, D. and {Steyn}, T.~J. and {Strong}, R. and {Struthers}, A. and {Stuart}, C. and {Sunnylall}, P. and {Swart}, P.~S. and {Taljaard}, B. and {Tasse}, C. and {Taylor}, G. and {Theron}, I.~P. and {Thondikulam}, V. and {Thorat}, K. and {Tiplady}, A. and {Toruvanda}, O. and {van Aardt}, J. and {van Balla}, T. and {van den Heever}, L. and {van der Byl}, A. and {van der Merwe}, C. and {van der Merwe}, P. and {van Niekerk}, P.~C. and {van Rooyen}, R. and {van Staden}, J.~P. and {van Tonder}, V. and {van Wyk}, R.},
        title = "{Revival of the Magnetar PSR J1622-4950: Observations with MeerKAT, Parkes, XMM-Newton, Swift, Chandra, and NuSTAR}",
      journal = {\apj},
         year = 2018,
        month = apr,
       volume = {856},
       number = {2},
          eid = {180},
        pages = {180},
          doi = {10.3847/1538-4357/aab35a},
archivePrefix = {arXiv},
       eprint = {1804.01933},
 primaryClass = {astro-ph.HE},
       adsurl = {https://ui.adsabs.harvard.edu/abs/2018ApJ...856..180C}
}

@ARTICLE{20231e,
       author = {{Lower}, Marcus E. and {Younes}, George and {Scholz}, Paul and {Camilo}, Fernando and {Dunn}, Liam and {Johnston}, Simon and {Enoto}, Teruaki and {Sarkissian}, John M. and {Reynolds}, John E. and {Palmer}, David M. and {Arzoumanian}, Zaven and {Baring}, Matthew G. and {Gendreau}, Keith and {G{\"o}{\u{g}}{\"u}{\textcommabelow s}}, Ersin and {Guillot}, Sebastien and {van der Horst}, Alexander J. and {Hu}, Chin-Ping and {Kouveliotou}, Chryssa and {Lin}, Lin and {Malacaria}, Christian and {Stewart}, Rachael and {Wadiasingh}, Zorawar},
        title = "{The 2022 High-energy Outburst and Radio Disappearing Act of the Magnetar 1E 1547.0-5408}",
      journal = {\apj},
         year = 2023,
        month = mar,
       volume = {945},
       number = {2},
          eid = {153},
        pages = {153},
          doi = {10.3847/1538-4357/acbc7c},
archivePrefix = {arXiv},
       eprint = {2302.07397},
 primaryClass = {astro-ph.HE},
       adsurl = {https://ui.adsabs.harvard.edu/abs/2023ApJ...945..153L}
}

@ARTICLE{20081e,
       author = {{Camilo}, F. and {Reynolds}, J. and {Johnston}, S. and {Halpern}, J.~P. and {Ransom}, S.~M.},
        title = "{The Magnetar 1E 1547.0-5408: Radio Spectrum, Polarimetry, and Timing}",
      journal = {\apj},
         year = 2008,
        month = may,
       volume = {679},
       number = {1},
        pages = {681-686},
          doi = {10.1086/587054},
archivePrefix = {arXiv},
       eprint = {0802.0494},
 primaryClass = {astro-ph},
       adsurl = {https://ui.adsabs.harvard.edu/abs/2008ApJ...679..681C}
}

@ARTICLE{2020sgr1,
       author = {{CHIME/FRB Collaboration} and {Andersen}, B.~C. and {Bandura}, K.~M. and {Bhardwaj}, M. and {Bij}, A. and {Boyce}, M.~M. and {Boyle}, P.~J. and {Brar}, C. and {Cassanelli}, T. and {Chawla}, P. and {Chen}, T. and {Cliche}, J.-F. and {Cook}, A. and {Cubranic}, D. and {Curtin}, A.~P. and {Denman}, N.~T. and {Dobbs}, M. and {Dong}, F.~Q. and {Fandino}, M. and {Fonseca}, E. and {Gaensler}, B.~M. and {Giri}, U. and {Good}, D.~C. and {Halpern}, M. and {Hill}, A.~S. and {Hinshaw}, G.~F. and {H{\"o}fer}, C. and {Josephy}, A. and {Kania}, J.~W. and {Kaspi}, V.~M. and {Landecker}, T.~L. and {Leung}, C. and {Li}, D.~Z. and {Lin}, H.-H. and {Masui}, K.~W. and {McKinven}, R. and {Mena-Parra}, J. and {Merryfield}, M. and {Meyers}, B.~W. and {Michilli}, D. and {Milutinovic}, N. and {Mirhosseini}, A. and {M{\"u}nchmeyer}, M. and {Naidu}, A. and {Newburgh}, L.~B. and {Ng}, C. and {Patel}, C. and {Pen}, U.-L. and {Pinsonneault-Marotte}, T. and {Pleunis}, Z. and {Quine}, B.~M. and {Rafiei-Ravandi}, M. and {Rahman}, M. and {Ransom}, S.~M. and {Renard}, A. and {Sanghavi}, P. and {Scholz}, P. and {Shaw}, J.~R. and {Shin}, K. and {Siegel}, S.~R. and {Singh}, S. and {Smegal}, R.~J. and {Smith}, K.~M. and {Stairs}, I.~H. and {Tan}, C.~M. and {Tendulkar}, S.~P. and {Tretyakov}, I. and {Vanderlinde}, K. and {Wang}, H. and {Wulf}, D. and {Zwaniga}, A.~V.},
        title = "{A bright millisecond-duration radio burst from a Galactic magnetar}",
      journal = {\nat},
         year = 2020,
        month = nov,
       volume = {587},
       number = {7832},
        pages = {54-58},
          doi = {10.1038/s41586-020-2863-y},
archivePrefix = {arXiv},
       eprint = {2005.10324},
 primaryClass = {astro-ph.HE},
       adsurl = {https://ui.adsabs.harvard.edu/abs/2020Natur.587...54C}
}

@ARTICLE{2021sgr2,
       author = {{Kirsten}, F. and {Snelders}, M.~P. and {Jenkins}, M. and {Nimmo}, K. and {van den Eijnden}, J. and {Hessels}, J.~W.~T. and {Gawro{\'n}ski}, M.~P. and {Yang}, J.},
        title = "{Detection of two bright radio bursts from magnetar SGR 1935 + 2154}",
      journal = {Nature Astronomy},
         year = 2021,
        month = apr,
       volume = {5},
        pages = {414-422},
          doi = {10.1038/s41550-020-01246-3},
archivePrefix = {arXiv},
       eprint = {2007.05101},
 primaryClass = {astro-ph.HE},
       adsurl = {https://ui.adsabs.harvard.edu/abs/2021NatAs...5..414K}
}

@ARTICLE{2020sgr3,
       author = {{Zhang}, C.~F. and {Jiang}, J.~C. and {Men}, Y.~P. and {Wang}, B.~J. and {Xu}, H. and {Xu}, J.~W. and {Niu}, C.~H. and {Zhou}, D.~J. and {Guan}, X. and {Han}, J.~L. and {Jiang}, P. and {Lee}, K.~J. and {Li}, D. and {Lin}, L. and {Niu}, J.~R. and {Wang}, P. and {Wang}, Z.~L. and {Xu}, R.~X. and {Yu}, W. and {Zhang}, B. and {Zhu}, W.~W.},
        title = "{A highly polarised radio burst detected from SGR 1935+2154 by FAST}",
      journal = {The Astronomer's Telegram},
         year = 2020,
        month = may,
       volume = {13699},
        pages = {1},
       adsurl = {https://ui.adsabs.harvard.edu/abs/2020ATel13699....1Z}
}

@ARTICLE{2019phase,
       author = {{Ilie}, C.~D. and {Johnston}, S. and {Weltevrede}, P.},
        title = "{Evidence for magnetospheric effects on the radiation of radio pulsars}",
      journal = {\mnras},
         year = 2019,
        month = feb,
       volume = {483},
       number = {2},
        pages = {2778-2794},
          doi = {10.1093/mnras/sty3315},
archivePrefix = {arXiv},
       eprint = {1811.12831},
 primaryClass = {astro-ph.HE},
       adsurl = {https://ui.adsabs.harvard.edu/abs/2019MNRAS.483.2778I}
}

@ARTICLE{2018gcp,
       author = {{Desvignes}, G. and {Eatough}, R.~P. and {Pen}, U.~L. and {Lee}, K.~J. and {Mao}, S.~A. and {Karuppusamy}, R. and {Schnitzeler}, D.~H.~F.~M. and {Falcke}, H. and {Kramer}, M. and {Wucknitz}, O. and {Spitler}, L.~G. and {Torne}, P. and {Liu}, K. and {Bower}, G.~C. and {Cognard}, I. and {Lyne}, A.~G. and {Stappers}, B.~W.},
        title = "{Large Magneto-ionic Variations toward the Galactic Center Magnetar, PSR J1745-2900}",
      journal = {\apjl},
         year = 2018,
        month = jan,
       volume = {852},
       number = {1},
          eid = {L12},
        pages = {L12},
          doi = {10.3847/2041-8213/aaa2f8},
archivePrefix = {arXiv},
       eprint = {1711.10323},
 primaryClass = {astro-ph.HE},
       adsurl = {https://ui.adsabs.harvard.edu/abs/2018ApJ...852L..12D}
}

@ARTICLE{2023wangpsr,
       author = {{Wang}, S.~Q. and {Wang}, J.~B. and {Li}, D.~Z. and {Yao}, J.~M. and {Manchester}, R.~N. and {Hobbs}, G. and {Wang}, N. and {Dai}, S. and {Xu}, H. and {Luo}, R. and {Feng}, Y. and {Wang}, W.~Y. and {Li}, D. and {Yu}, Y.~W. and {Du}, Z.~X. and {Niu}, C.~H. and {Zhang}, S.~B. and {Zhang}, C.~M.},
        title = "{Change of Rotation Measure during the Eclipse of a Black Widow PSR J2051-0827}",
      journal = {\apj},
         year = 2023,
        month = sep,
       volume = {955},
       number = {1},
          eid = {36},
        pages = {36},
          doi = {10.3847/1538-4357/acea81},
archivePrefix = {arXiv},
       eprint = {2307.13198},
 primaryClass = {astro-ph.HE},
       adsurl = {https://ui.adsabs.harvard.edu/abs/2023ApJ...955...36W}
}

@ARTICLE{2005psr1269,
       author = {{Johnston}, Simon and {Ball}, Lewis and {Wang}, N. and {Manchester}, R.~N.},
        title = "{Radio observations of PSR B1259-63 through the 2004 periastron passage}",
      journal = {\mnras},
         year = 2005,
        month = apr,
       volume = {358},
       number = {3},
        pages = {1069-1075},
          doi = {10.1111/j.1365-2966.2005.08854.x},
archivePrefix = {arXiv},
       eprint = {astro-ph/0501660},
 primaryClass = {astro-ph},
       adsurl = {https://ui.adsabs.harvard.edu/abs/2005MNRAS.358.1069J}
}

@ARTICLE{2022Nat76s,
       author = {{Caleb}, Manisha and {Heywood}, Ian and {Rajwade}, Kaustubh and {Malenta}, Mateusz and {Stappers}, Benjamin Willem and {Barr}, Ewan and {Chen}, Weiwei and {Morello}, Vincent and {Sanidas}, Sotiris and {van den Eijnden}, Jakob and {Kramer}, Michael and {Buckley}, David and {Brink}, Jaco and {Motta}, Sara Elisa and {Woudt}, Patrick and {Weltevrede}, Patrick and {Jankowski}, Fabian and {Surnis}, Mayuresh and {Buchner}, Sarah and {Bezuidenhout}, Mechiel Christiaan and {Driessen}, Laura Nicole and {Fender}, Rob},
        title = "{Discovery of a radio-emitting neutron star with an ultra-long spin period of 76 s}",
      journal = {\nastro},
         year = 2022,
        month = may,
       volume = {6},
        pages = {828-836},
          doi = {10.1038/s41550-022-01688-x},
archivePrefix = {arXiv},
       eprint = {2206.01346},
 primaryClass = {astro-ph.HE},
       adsurl = {https://ui.adsabs.harvard.edu/abs/2022NatAs...6..828C}
}

@ARTICLE{2022Nat18min,
       author = {{Hurley-Walker}, N. and {Zhang}, X. and {Bahramian}, A. and {McSweeney}, S.~J. and {O'Doherty}, T.~N. and {Hancock}, P.~J. and {Morgan}, J.~S. and {Anderson}, G.~E. and {Heald}, G.~H. and {Galvin}, T.~J.},
        title = "{A radio transient with unusually slow periodic emission}",
      journal = {\nat},
         year = 2022,
        month = jan,
       volume = {601},
       number = {7894},
        pages = {526-530},
          doi = {10.1038/s41586-021-04272-x},
archivePrefix = {arXiv},
       eprint = {2503.08033},
 primaryClass = {astro-ph.HE},
       adsurl = {https://ui.adsabs.harvard.edu/abs/2022Natur.601..526H}
}

@ARTICLE{li202444,
       author = {{Li}, Di and {Yuan}, Mao and {Wu}, Lin and {Yan}, Jingye and {Lv}, Xuning and {Tsai}, Chao-Wei and {Wang}, Pei and {Zhu}, WeiWei and {Deng}, Li and {Lan}, Ailan and {Xu}, Renxin and {Chen}, Xianglei and {Meng}, Lingqi and {Li}, Jian and {Li}, Xiangdong and {Zhou}, Ping and {Yang}, Haoran and {Xue}, Mengyao and {Lu}, Jiguang and {Miao}, Chenchen and {Wang}, Weiyang and {Niu}, Jiarui and {Fang}, Ziyao and {Fu}, Qiuyang and {Feng}, Yi and {Zhang}, Peijin and {Jiang}, Jinchen and {Miao}, Xueli and {Chen}, Yu and {Sun}, Lingchen and {Yang}, Yang and {Deng}, Xiang and {Dai}, Shi and {Chen}, Xue and {Yao}, Jumei and {Liu}, Yujie and {Li}, Changheng and {Zhang}, Minglu and {Yang}, Yiwen and {Zhou}, Yucheng and {Zhou}, Yi-Yi and {Zhang}, Yongkun and {Niu}, Chenhui and {Zhao}, Rushuang and {Zhang}, Lei and {Peng}, Bo and {Wu}, Ji and {Wang}, Chi},
        title = "{A 44-minute periodic radio transient in a supernova remnant.}",
note        = {Preprint at http://arXiv.org/abs/2411.15739 (2024)},
}

@article{men2026detection,
  author        = {Men, Yunpeng and Barr, Ewan and Qu, Yuanhong and Horvath, Csanad and Jiang, Jinchen and Desvignes, Gregory and Hurley-Walker, Natasha and Kramer, Michael and Luo, Rui and McSweeney, Samuel J. and Wu, Jason},
  title         = {Detection of Cyclotron Absorption in the Radio Emission of GPM 1839-10.},
  note        = {Preprint at http://arXiv.org/abs/2602.13742 (2026)}
}

@ARTICLE{2024meerkat,
       author = {{Keith}, M.~J. and {Johnston}, S. and {Karastergiou}, A. and {Weltevrede}, P. and {Lower}, M.~E. and {Basu}, A. and {Posselt}, B. and {Oswald}, L.~S. and {Parthasarathy}, A. and {Cameron}, A.~D. and {Serylak}, M. and {Buchner}, S.},
        title = "{The Thousand-Pulsar-Array programme on MeerKAT - XIII. Timing, flux density, rotation measure, and dispersion measure time series of 597 pulsars}",
      journal = {\mnras},
         year = 2024,
        month = may,
       volume = {530},
       number = {2},
        pages = {1581-1591},
          doi = {10.1093/mnras/stae937},
archivePrefix = {arXiv},
       eprint = {2404.02051},
 primaryClass = {astro-ph.HE},
       adsurl = {https://ui.adsabs.harvard.edu/abs/2024MNRAS.530.1581K}
}

@ARTICLE{2011mspsr,
       author = {{Yan}, W.~M. and {Manchester}, R.~N. and {Hobbs}, G. and {van Straten}, W. and {Reynolds}, J.~E. and {Wang}, N. and {Bailes}, M. and {Bhat}, N.~D.~R. and {Burke-Spolaor}, S. and {Champion}, D.~J. and {Chaudhary}, A. and {Coles}, W.~A. and {Hotan}, A.~W. and {Khoo}, J. and {Oslowski}, S. and {Sarkissian}, J.~M. and {Yardley}, D.~R.~B.},
        title = "{Rotation measure variations for 20 millisecond pulsars}",
      journal = {\apss},
         year = 2011,
        month = oct,
       volume = {335},
       number = {2},
        pages = {485-498},
          doi = {10.1007/s10509-011-0756-0},
archivePrefix = {arXiv},
       eprint = {1105.4213},
 primaryClass = {astro-ph.SR},
       adsurl = {https://ui.adsabs.harvard.edu/abs/2011Ap&SS.335..485Y}
}

@ARTICLE{2002bpsr,
       author = {{Connors}, T.~W. and {Johnston}, S. and {Manchester}, R.~N. and {McConnell}, D.},
        title = "{The 2000 periastron passage of PSR B1259-63}",
      journal = {\mnras},
         year = 2002,
        month = nov,
       volume = {336},
       number = {4},
        pages = {1201-1208},
          doi = {10.1046/j.1365-8711.2002.05850.x},
archivePrefix = {arXiv},
       eprint = {astro-ph/0207302},
 primaryClass = {astro-ph},
       adsurl = {https://ui.adsabs.harvard.edu/abs/2002MNRAS.336.1201C}
}

@ARTICLE{2021frb121102,
       author = {{Hilmarsson}, G.~H. and {Michilli}, D. and {Spitler}, L.~G. and {Wharton}, R.~S. and {Demorest}, P. and {Desvignes}, G. and {Gourdji}, K. and {Hackstein}, S. and {Hessels}, J.~W.~T. and {Nimmo}, K. and {Seymour}, A.~D. and {Kramer}, M. and {Mckinven}, R.},
        title = "{Rotation Measure Evolution of the Repeating Fast Radio Burst Source FRB 121102}",
      journal = {\apjl},
         year = 2021,
        month = feb,
       volume = {908},
       number = {1},
          eid = {L10},
        pages = {L10},
          doi = {10.3847/2041-8213/abdec0},
archivePrefix = {arXiv},
       eprint = {2009.12135},
 primaryClass = {astro-ph.HE},
       adsurl = {https://ui.adsabs.harvard.edu/abs/2021ApJ...908L..10H}
}

@ARTICLE{2023chimefrb,
       author = {{Mckinven}, R. and {Gaensler}, B.~M. and {Michilli}, D. and {Masui}, K. and {Kaspi}, V.~M. and {Su}, J. and {Bhardwaj}, M. and {Cassanelli}, T. and {Chawla}, P. and {Dong}, F. Adam and {Fonseca}, E. and {Leung}, C. and {Li}, D.~Z. and {Ng}, C. and {Patel}, C. and {Pearlman}, A.~B. and {Petroff}, E. and {Pleunis}, Z. and {Rafiei-Ravandi}, M. and {Rahman}, M. and {Sand}, K.~R. and {Shin}, K. and {Stairs}, I.~H. and {Tendulkar}, S.},
        title = "{Revealing the Dynamic Magnetoionic Environments of Repeating Fast Radio Burst Sources through Multiyear Polarimetric Monitoring with CHIME/FRB}",
      journal = {\apj},
         year = 2023,
        month = jul,
       volume = {951},
       number = {1},
          eid = {82},
        pages = {82},
          doi = {10.3847/1538-4357/acd188},
archivePrefix = {arXiv},
       eprint = {2302.08386},
 primaryClass = {astro-ph.HE},
       adsurl = {https://ui.adsabs.harvard.edu/abs/2023ApJ...951...82M}
}

@ARTICLE{2025Sfastrepeaterrm,
       author = {{Feng}, Yi and {Zhang}, Yong-Kun and {Xie}, Jintao and {Yang}, Yuan-Pei and {Qu}, Yuanhong and {Zhou}, Dengke and {Li}, Di and {Zhang}, Bing and {Zhu}, Weiwei and {Lu}, Wenbin and {Xu}, Jiaying and {Miao}, Chenchen and {Tian}, Shiyan and {Wang}, Pei and {Yao}, Ju-Mei and {Niu}, Chen-Hui and {Niu}, Jiarui and {Xu}, Heng and {Jiang}, Jinchen and {Zhou}, Dejiang and {Liu}, Zenan and {Tsai}, Chao-Wei and {Dai}, Zigao and {Wu}, Xuefeng and {Wang}, Fayin and {Han}, Jinlin and {Lee}, Kejia and {Xu}, Renxin and {Huang}, Yongfeng and {Zou}, Yuanchuan and {Cao}, Jinhuang and {Chen}, Xianglei and {Fang}, Jianhua and {Li}, Dongzi and {Li}, Ye and {Lu}, Wanjin and {Luo}, Jiawei and {Luo}, Jintao and {Luo}, Rui and {Lyu}, Fen and {Wang}, Bojun and {Wang}, Weiyang and {Wu}, Qin and {Xue}, Mengyao and {Xiao}, Di and {Yu}, Wenfei and {Yuan}, Jianping and {Zhang}, Chunfeng and {Zhang}, Junshuo and {Zhang}, Lei and {Zhang}, Songbo and {Zhao}, Rushuang and {Zhu}, Yuhao},
        title = "{Multi-year polarimetric monitoring of four CHIME-discovered repeating fast radio bursts with FAST}",
      journal = {Science China Physics, Mechanics, and Astronomy},
         year = 2025,
        month = aug,
       volume = {68},
       number = {8},
          eid = {289511},
        pages = {289511},
          doi = {10.1007/s11433-024-2668-5},
archivePrefix = {arXiv},
       eprint = {2507.02355},
 primaryClass = {astro-ph.HE},
       adsurl = {https://ui.adsabs.harvard.edu/abs/2025SCPMA..6889511F}
}

@ARTICLE{2023frb,
       author = {{Anna-Thomas}, Reshma and {Connor}, Liam and {Dai}, Shi and {Feng}, Yi and {Burke-Spolaor}, Sarah and {Beniamini}, Paz and {Yang}, Yuan-Pei and {Zhang}, Yong-Kun and {Aggarwal}, Kshitij and {Law}, Casey J. and {Li}, Di and {Niu}, Chenhui and {Chatterjee}, Shami and {Cruces}, Marilyn and {Duan}, Ran and {Filipovic}, Miroslav D. and {Hobbs}, George and {Lynch}, Ryan S. and {Miao}, Chenchen and {Niu}, Jiarui and {Ocker}, Stella K. and {Tsai}, Chao-Wei and {Wang}, Pei and {Xue}, Mengyao and {Yao}, Ju-Mei and {Yu}, Wenfei and {Zhang}, Bing and {Zhang}, Lei and {Zhu}, Shiqiang and {Zhu}, Weiwei},
        title = "{Magnetic field reversal in the turbulent environment around a repeating fast radio burst}",
      journal = {Science},
         year = 2023,
        month = may,
       volume = {380},
       number = {6645},
        pages = {599-603},
          doi = {10.1126/science.abo6526},
archivePrefix = {arXiv},
       eprint = {2202.11112},
 primaryClass = {astro-ph.HE},
       adsurl = {https://ui.adsabs.harvard.edu/abs/2023Sci...380..599A}
}

@ARTICLE{1998cp,
       author = {{Han}, J.~L. and {Manchester}, R.~N. and {Xu}, R.~X. and {Qiao}, G.~J.},
        title = "{Circular polarization in pulsar integrated profiles}",
      journal = {\mnras},
         year = 1998,
        month = oct,
       volume = {300},
       number = {2},
        pages = {373-387},
          doi = {10.1046/j.1365-8711.1998.01869.x},
archivePrefix = {arXiv},
       eprint = {astro-ph/9806021},
 primaryClass = {astro-ph},
       adsurl = {https://ui.adsabs.harvard.edu/abs/1998MNRAS.300..373H}
}

\section*{Data availability statement}
FAST observational data are made publicly available one year after the observations. All public FAST data are available from the FAST user website, \url{http://fast.bao.ac.cn}. The Insight-HXMT data used in this work are available from HXMT website, \url{http://hxmten.ihep.ac.cn}.

\section*{Code availability}
PRESTO (\url{https://github.com/scottransom/presto})\\
astropy  (\url{https://www.astropy.org/})\\
DM-power (\url{https://github.com/hsiuhsil/DM-power})\\
RM-Tools (\url{https://github.com/CIRADA-Tools/RM-Tools})\\

\section*{Acknowledgements}
This work is supported by the NSFC (12133007) and National Key Research and Development Program of China (Grants No. 2023YFA1607901 and 2021YFA0718503). This work made use of the data from FAST (Five-hundred-meter Aperture Spherical radio Telescope). FAST is a Chinese national mega-science facility, operated by National Astronomical Observatories, Chinese Academy of Sciences.

\section*{Author contributions statement}
W.W. as the PI of the FAST observations led the project, data analysis, and wrote the paper. B.L. made the data analysis and wrote the part of the paper, Y.J. made X-ray data analysis and wrote the part of the methods, S.Q., C.G. and P.W. joined the FAST data analysis, S.J. made the X-ray observations, J.L. and W.W. made the science interpretation. All authors have reviewed the present results and manuscript.

\section*{Competing interests}
The authors declare no competing interests.

\clearpage

\section*{Extended Data}
\renewcommand\thefigure{\arabic{figure}} 
\setcounter{figure}{0}
\renewcommand{\figurename}{Extended Data Figure}
\renewcommand{\thetable}{\arabic{table}}
\setcounter{table}{0}
\renewcommand{\tablename}{Extended Data Table}

\begin{table}[htbp]
\centering
\caption{Observation information}
\label{tab:obs}
\begin{tabular}{lllllll}
\toprule
UTC & Mode & Length (s) & $T_{\rm samp}$ (s) & $N_{\rm chan}$& RA&DEC \\
\midrule
2025-08-30 06:00:00           & Tracking & 6000 & 0.000098304 & 4096 &06:30:38.40&	+25:26:23.0 \\
2026-02-06 21:45:00           & Tracking & 5520 & 0.000098304 & 4096 &06:30:38.40&  +25:26:23.0\\
2026-02-27 18:20:00$^{a}$     & Tracking & 6000 & 0.000098304 & 4096 &06:30:38.40&	+25:26:23.0\\
2026-03-21 18:25:00           & Tracking & 3420 & 0.000098304 & 4096 &06:30:38.40&	+25:26:23.0\\
\bottomrule
\end{tabular}

\begin{flushleft}
\footnotesize{$^{a}$ This observation is subject to strong interference. }
\end{flushleft}
\end{table}

\begin{table}[htbp]
\centering
\caption{Pulse properties. LP is measured from the single pulse with the highest degree of linear polarization, while CP is measured from the single pulse showing the clearest structure in the Stokes $V$ dynamic spectrum. A dash ("-") indicates that no reliable linear or circular polarization could be identified.}
\label{tab:info}
\begin{tabular}{llllll}
\toprule
$\rm MJD_{obs}$ & TOA (s)         & $\rm S_{peak}$(mJy) &$\rm LP$ &   CP  & $\rm RM(rad\ m^{-2})$\\
\midrule
60916.91666667  &3511.58-3511.80  &  $14\pm2$  & -       & -                 &    -  \\
60916.91666667  &3929.47-3930.55  &  $50\pm3$  &$36.73\pm3.50\%$ & $-19.64\pm3.45\%$ &  $-1386.87^{+9.72}_{-10.15}$    \\
60916.91666667  &4347.79-4348.38  &  $34\pm2$  &$55.87\pm3.4\%$ & $+43.39\pm7.4\%$   &   $-1337.35^{+3.72}_{-3.73}$ \\
60916.91666667  &5186.51-5186.77  &  $14\pm2$  &  -      & $+34.62\pm7.74\%$  &   -  \\
60916.91666667  &5605.47-5605.49  &  $11\pm2$  & -       & $+31.17\pm10.38\%$ &   -  \\
\\

61077.57291667  &358.78-360.86    &  $16\pm2$  &  -      &    -              &    -  \\
61077.57291667  &2857.21-2857.60  &  $22\pm2$  &  -      & -                 &  -    \\
61077.57291667  &5391.18-5394.30  &  $16\pm2$  & -       & $+3.38\pm5.6\%$    &  -    \\
\\

61098.43055556  &127.9-128.42     &  $15\pm5$  &  -      &  -                &     -  \\ 
61098.43055556  &2643.78-2644.47  &  $62\pm3$  & $19.50\pm1.6\%$& $+20.68\pm1.6\%$  &  $-1092.44^{+3.04}_{-3.37}$  \\ 
61098.43055556  &3063.36-3063.52  &  $8\pm2$   &    -    &          -        &  -     \\ 
61098.43055556  &3483.38-3483.57  &  $13\pm3$  &     -    &         -         &   -    \\ 
61098.43055556  &4322.97-4323.05  &  $14\pm3$  &      -   &  -                &   -    \\ 
61098.43055556  &5162.28-5162.79  &  $30\pm2$  & $21.08\pm2.44\%$& $+26.42\pm2.92\%$  &    $-1628.87^{+3.64}_{-2.86}$   \\ 
61098.43055556  &5581.55-5582.04  &  $16\pm2$  &      -   & $+71.62\pm2.01\%$  &   -    \\ 
\\

61120.43402778  &336.55-337.22 & $27\pm2$ & $32.67\pm3.55\%$  & $+20.18\pm3.44\%$ & $-487.14^{+7.7}_{-5.53}$  \\
\bottomrule
\end{tabular}
\begin{flushleft}
\end{flushleft}
\end{table}

\begin{figure}
    \centering
    \includegraphics[width=0.5\linewidth]{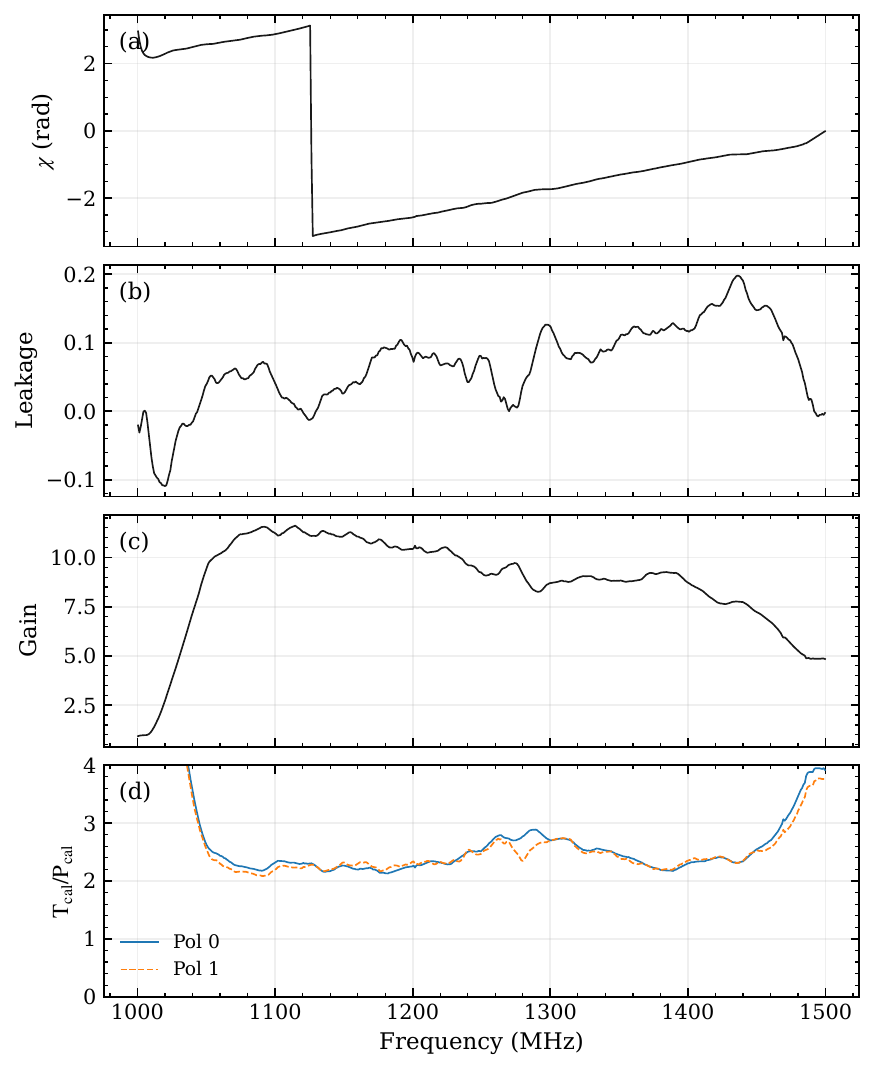}
    \caption{Calibration results of the observation on 2025 August 30. From top to bottom: phase delay, leakage, normalization factor, and the ratio between temperature and raw data intensity. }
    \label{fig:calres}
\end{figure}

\begin{figure*}[htbp]
    \centering
    \begin{subfigure}[b]{0.45\textwidth}
        \centering
        \includegraphics[width=\textwidth]{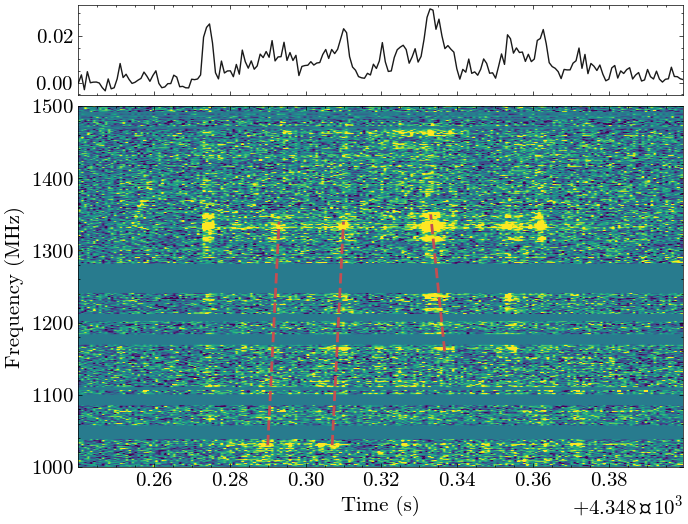}
        \caption{}
    \end{subfigure}
    \hfill
    \begin{subfigure}[b]{0.45\textwidth}
        \centering
        \includegraphics[width=\textwidth]{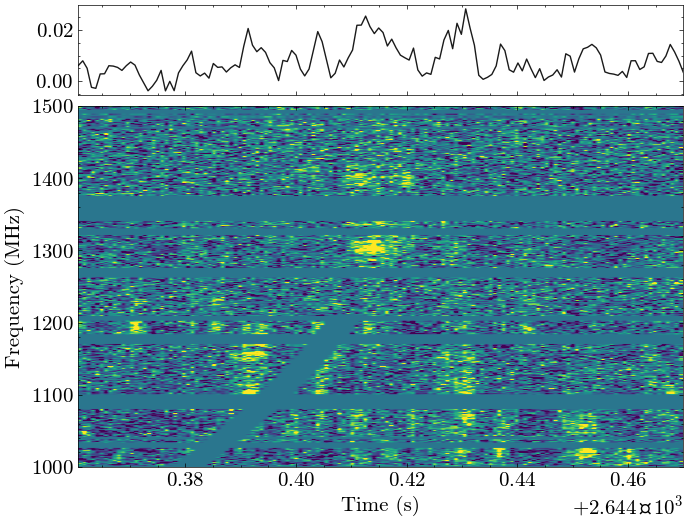}
        \caption{}
    \end{subfigure}
    \caption{
    Complex spectral morphology of CHIME J0630$+$25. In each panel, the upper and lower sub-panels show the Stokes $I$ light curve and dynamic spectrum, respectively. The dashed lines in \textbf{(A)} mark representative apparent frequency-drifting structures for the intra-pulse drifts. \textbf{(B)} shows the representative burst where inter-pulse drifts are present.
    }
    \label{fig:drift}
\end{figure*}

\begin{figure}
    \centering
    \includegraphics[width=\linewidth]{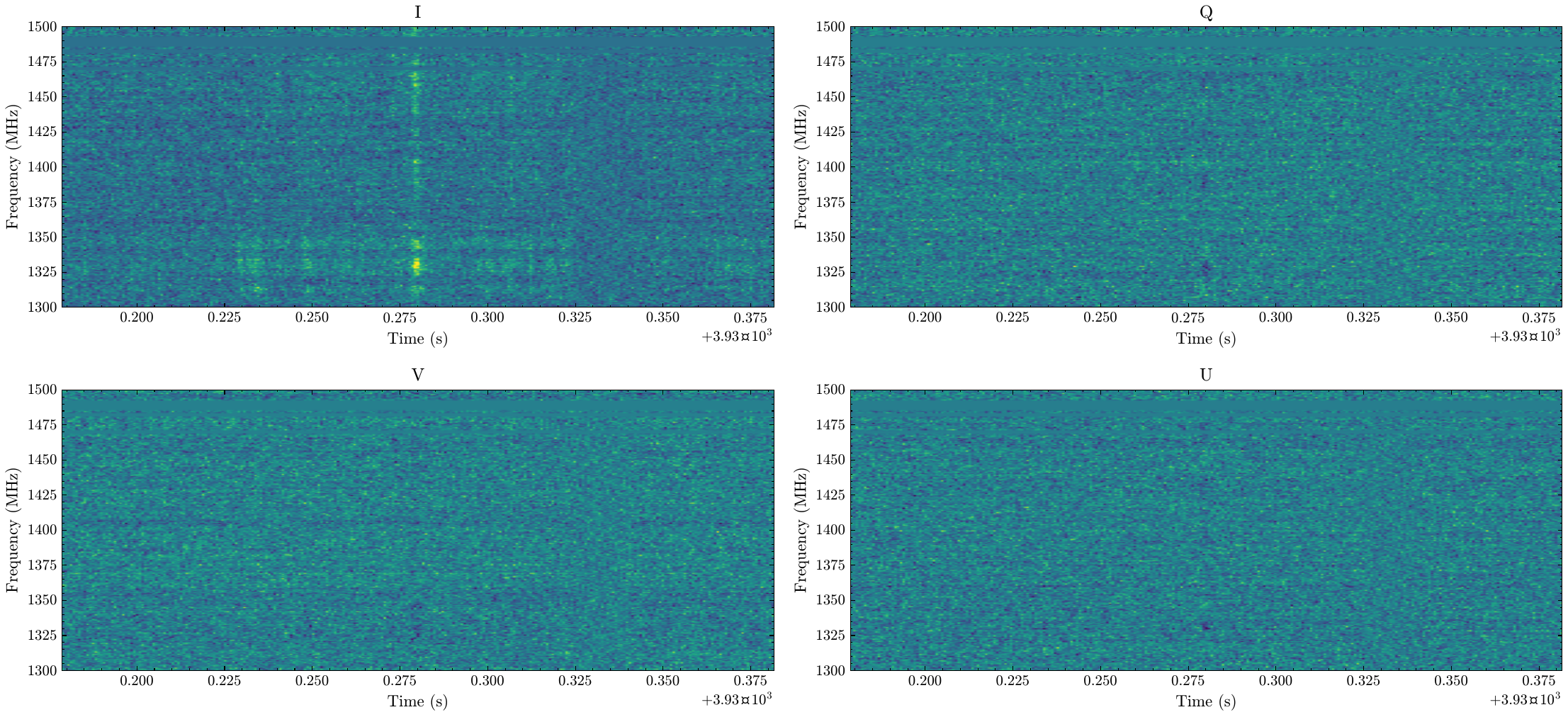}
    \caption{An example of the dynamic spectra of the Stokes parameters for a detected pulse exhibiting linear polarization (LP), observed at 3930.28 s on 2025 August 30. The upper-left, upper-right, lower-left, and lower-right panels present Stokes $I$, $Q$, $V$, and $U$, respectively. }
    \label{fig:IQUV}
\end{figure}

\begin{figure*}[htbp]
    \centering
    \begin{subfigure}[b]{0.32\textwidth}
        \centering
        \includegraphics[width=\textwidth]{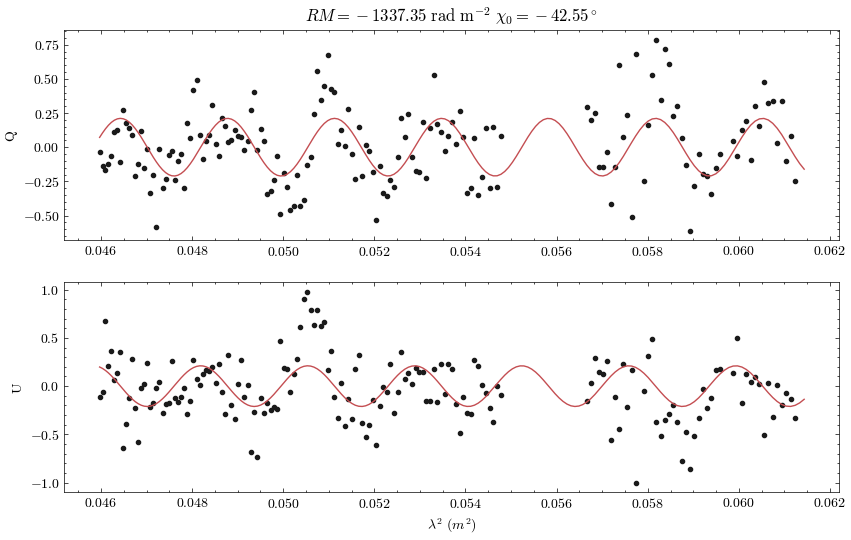}
        \caption{}
    \end{subfigure}
    \hfill
    \begin{subfigure}[b]{0.32\textwidth}
        \centering
        \includegraphics[width=\textwidth]{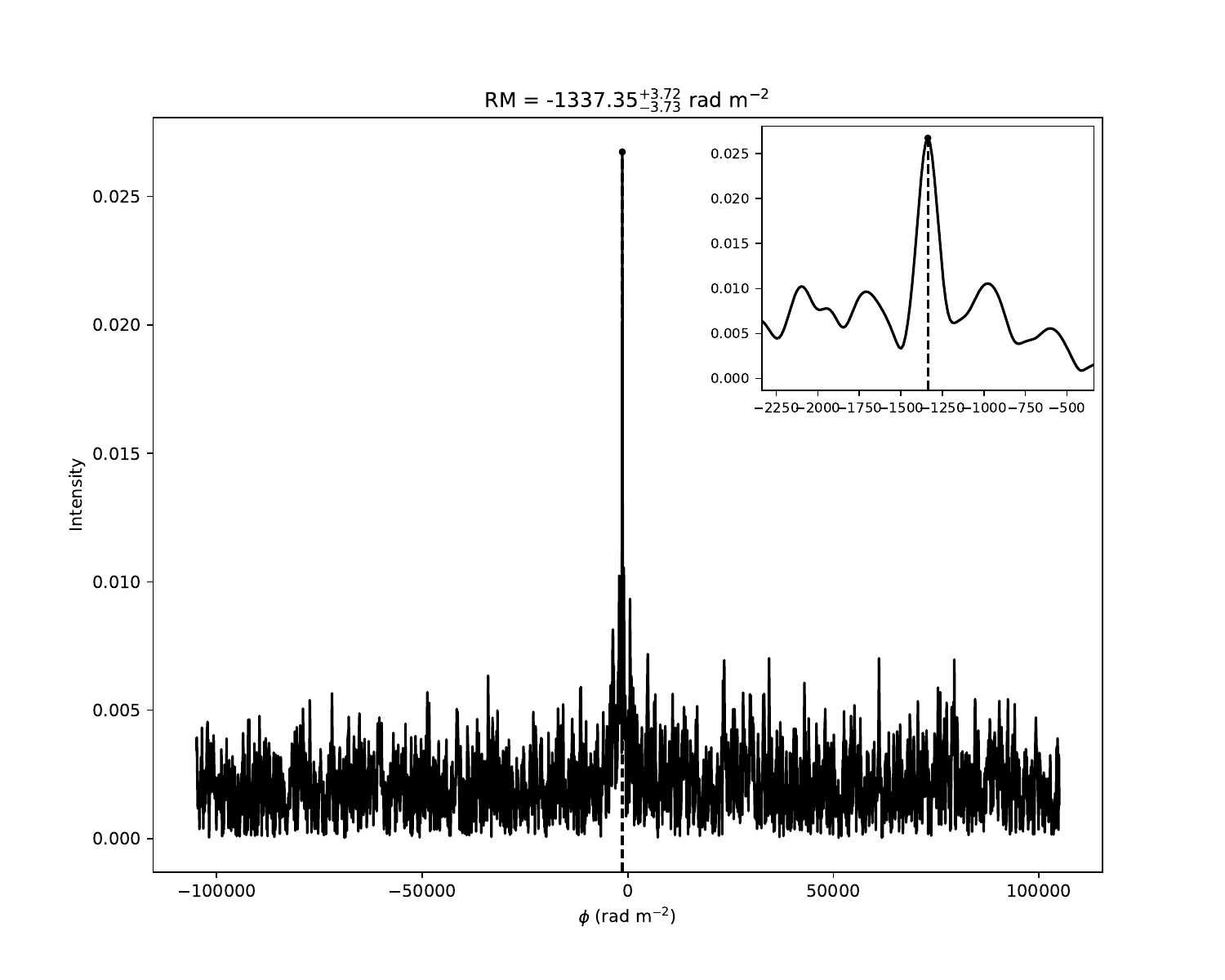}
        \caption{}
    \end{subfigure}
    \hfill
    \begin{subfigure}[b]{0.32\textwidth}
        \centering
        \includegraphics[width=\textwidth]{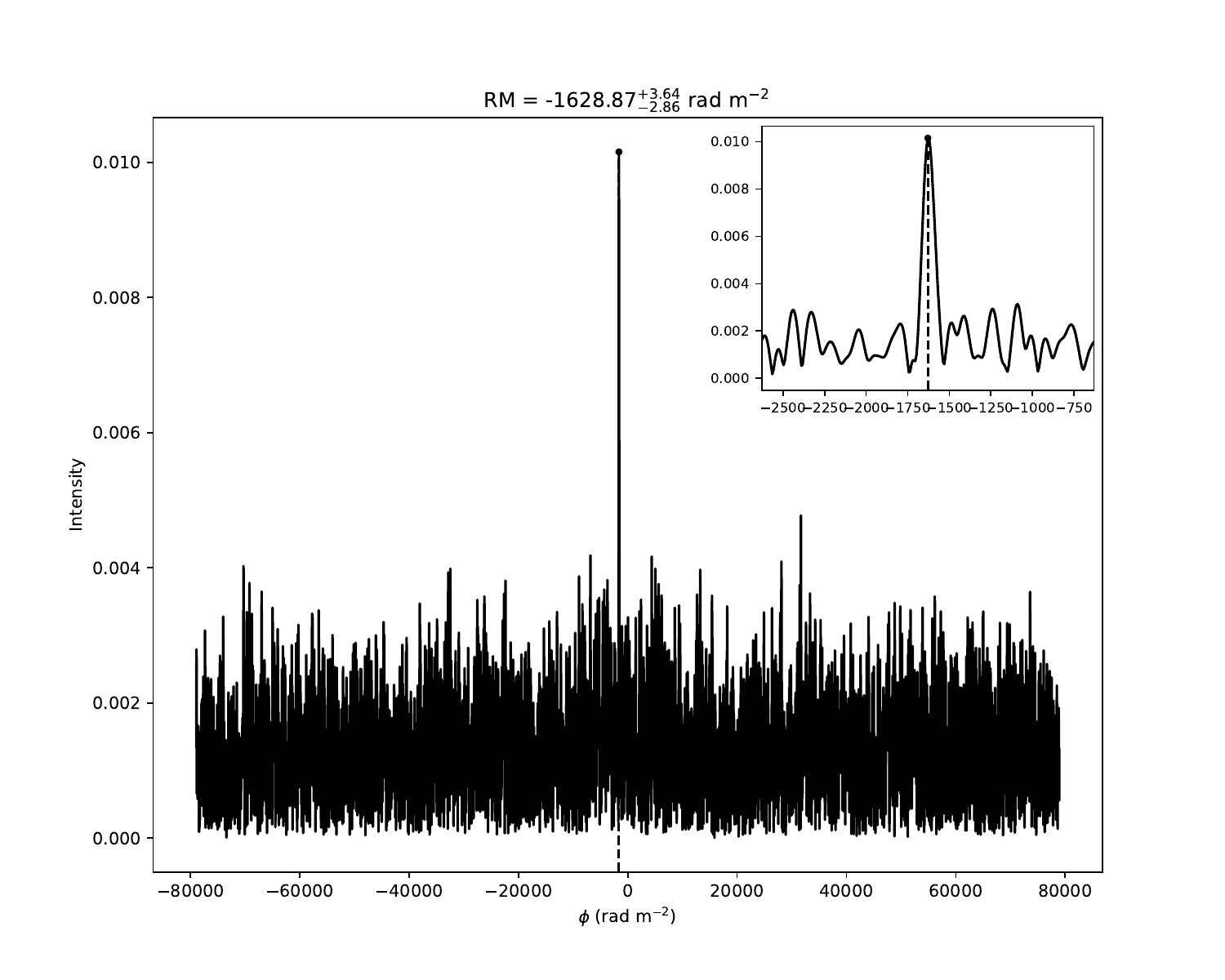}
        \caption{}
    \end{subfigure}

    \vspace{0.2cm}

    \begin{subfigure}[b]{0.32\textwidth}
        \centering
        \includegraphics[width=\textwidth]{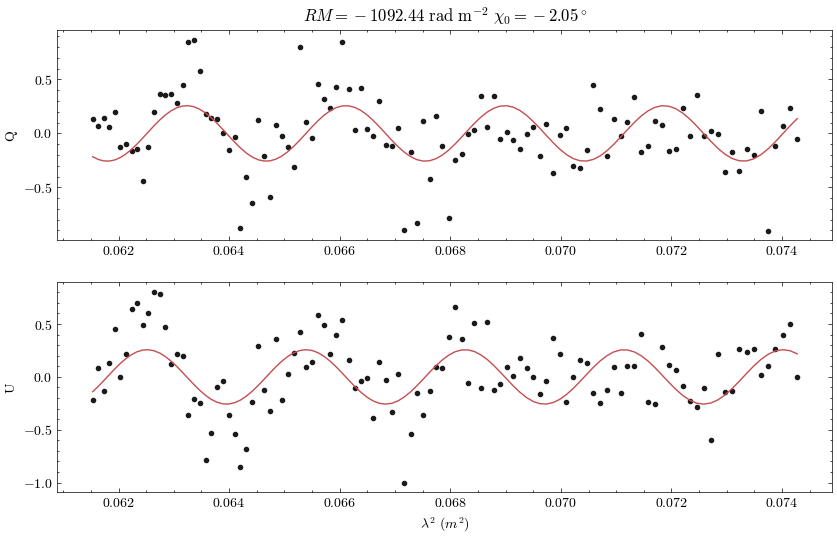}
        \caption{}
    \end{subfigure}
    \hfill
    \begin{subfigure}[b]{0.32\textwidth}
        \centering
        \includegraphics[width=\textwidth]{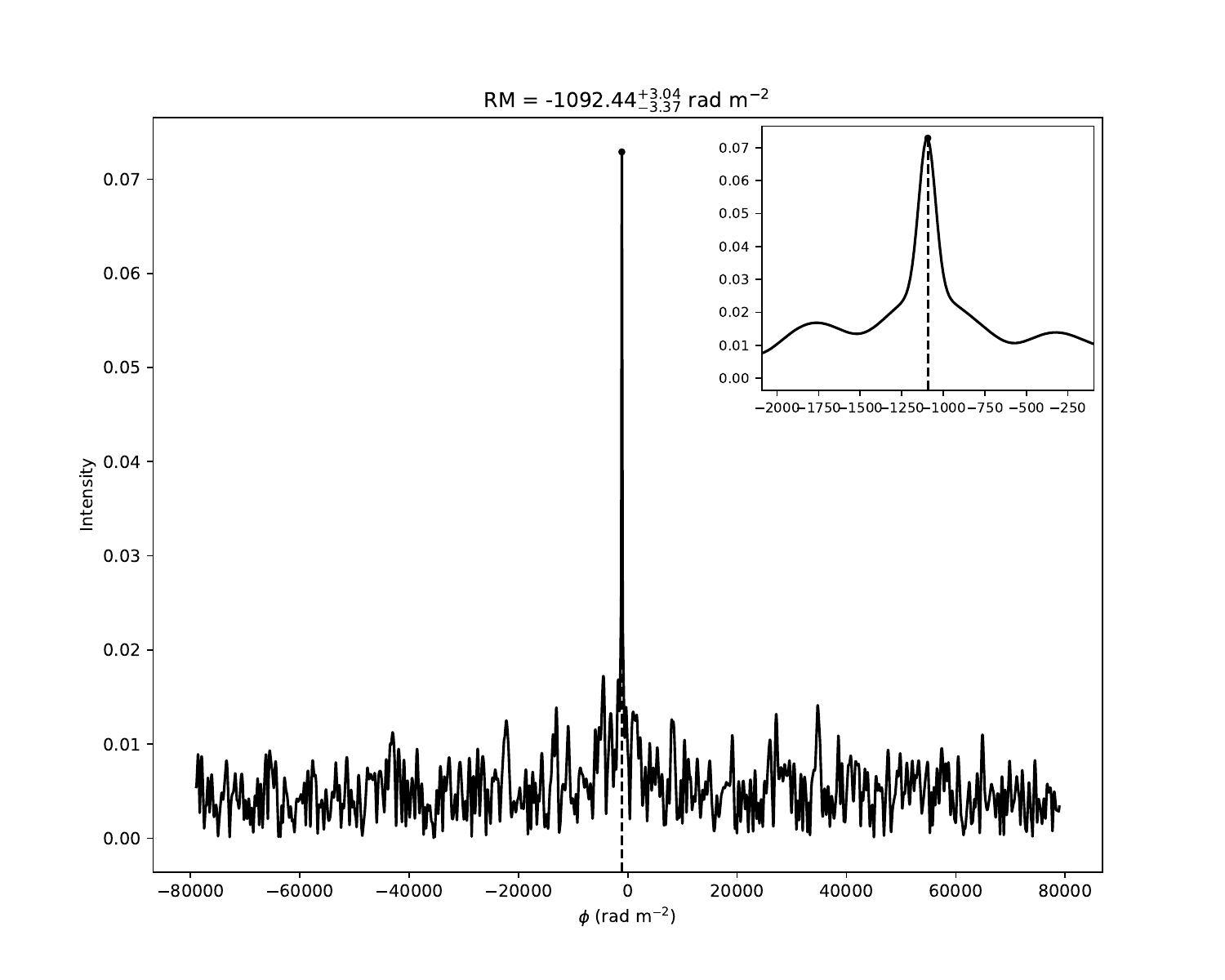}
        \caption{}
    \end{subfigure}
    \hfill
    \begin{subfigure}[b]{0.32\textwidth}
        \centering
        \includegraphics[width=\textwidth]{fig/260227/5162.642-5162.65-dm22_FDFclean.pdf}
        \caption{}
    \end{subfigure}

    \caption{The RM calculation results. \textbf{(A)} and \textbf{(D)} exhibit the fitting results of the Stokes $Q$ and $U$ using the RM obtained through RM synthesis with the RM and fitted intrinsic PA listed in its title. \textbf{(B)}, \textbf{(C)}, \textbf{(E)} and \textbf{(F)} show the RM results from RM synthesis after RM clean. A zoomed-in view is shown in the upper-right corner of each panel. }
    \label{fig:rmcal_all}
\end{figure*}

\begin{figure*}[htbp]
    \centering

    \begin{minipage}[c]{0.38\textwidth}
        \centering
        
        \begin{subfigure}[b]{\textwidth}
            \centering
            \includegraphics[width=\textwidth]{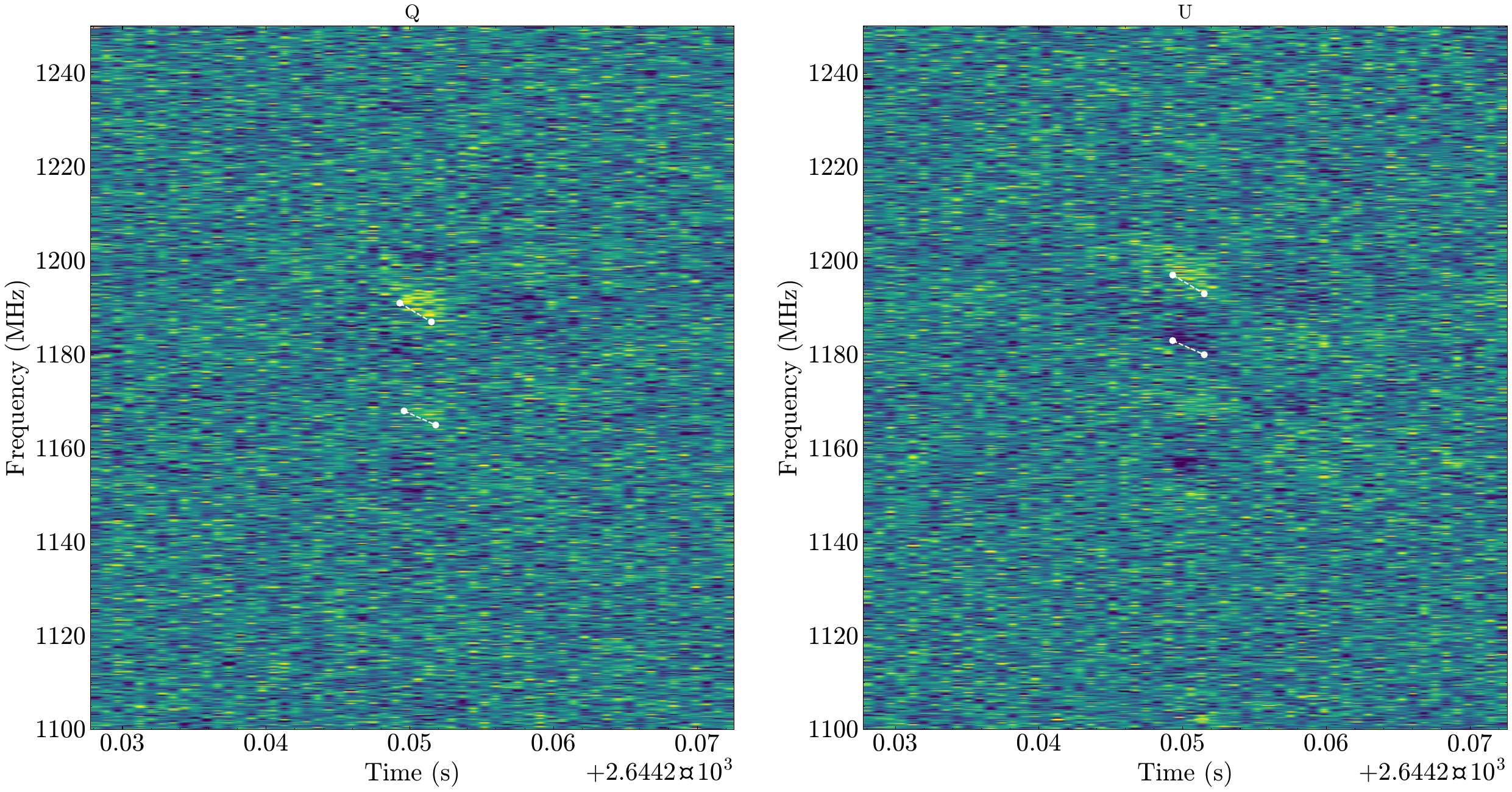}
        \end{subfigure}
        
        \vspace{0.3cm}
        
        \begin{subfigure}[b]{\textwidth}
            \centering
            \includegraphics[width=\textwidth]{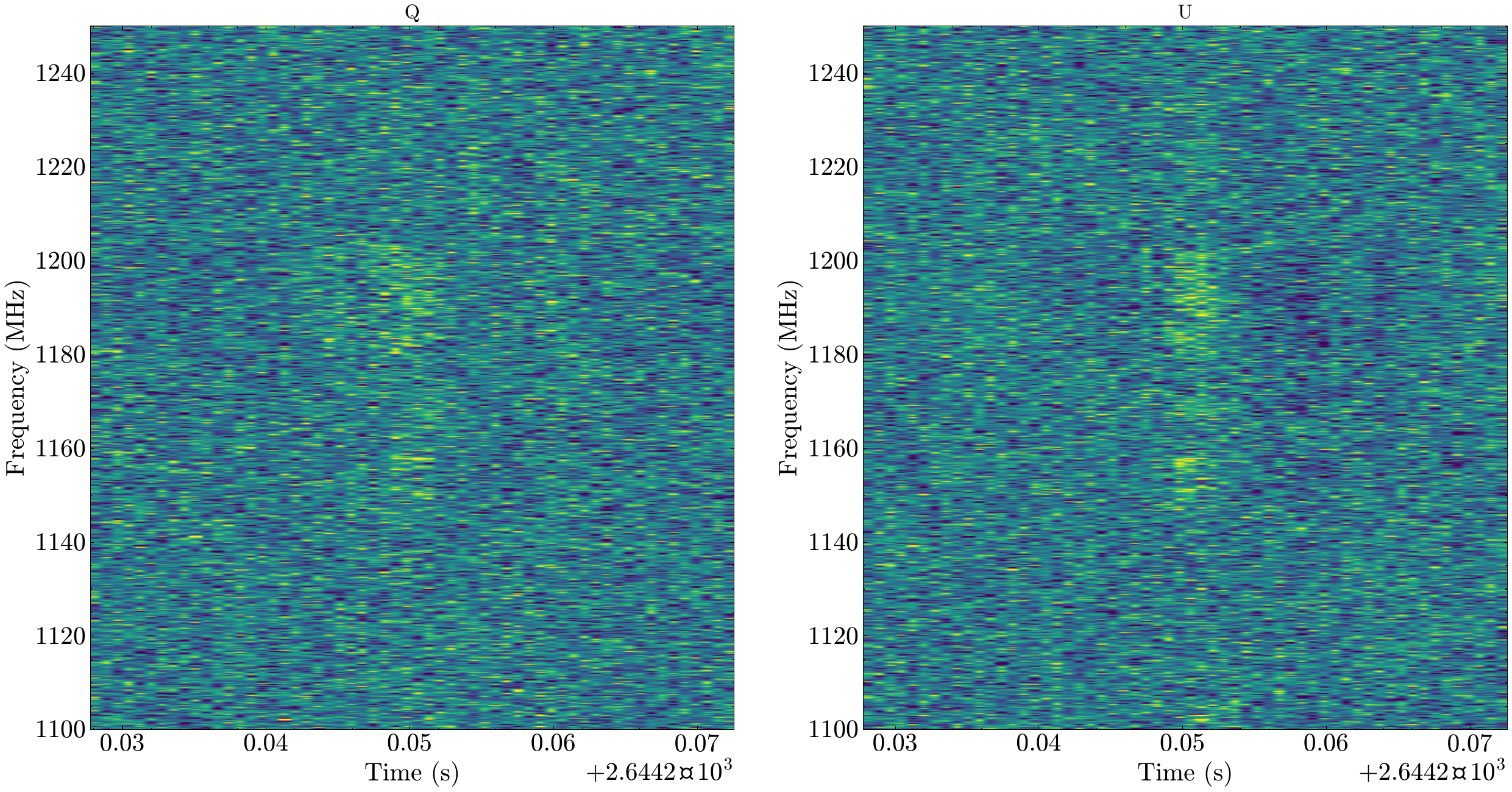}
        \end{subfigure}
    \end{minipage}
    \hfill
    \begin{minipage}[c]{0.58\textwidth}
        \centering
        \begin{subfigure}[b]{\textwidth}
            \centering
            \includegraphics[width=\textwidth,height=0.78\textheight,keepaspectratio]{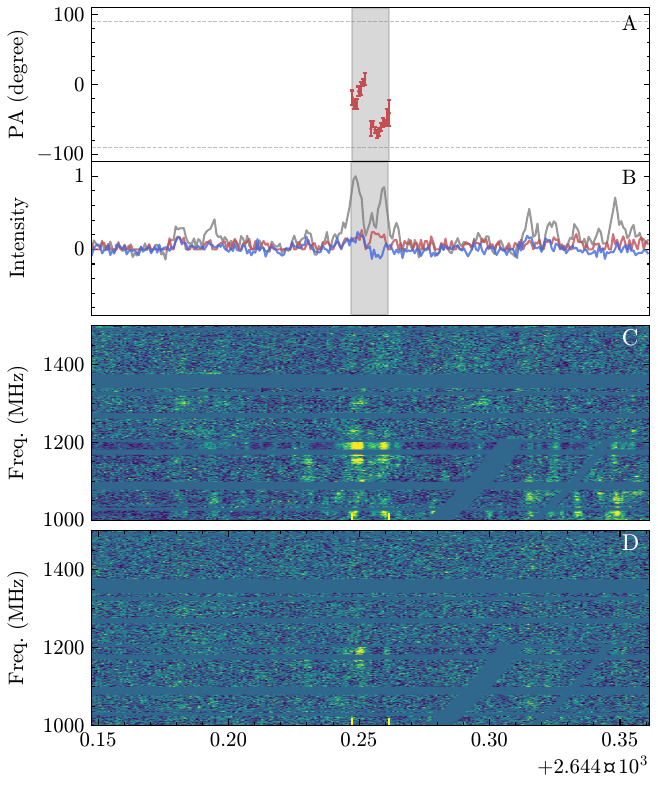}
        \end{subfigure}
    \end{minipage}

    \caption{The dynamic spectra of Stokes parameters. The left panels show the Stokes $Q$ and $U$ before and after Faraday derotation using $\rm RM=-1092.44\ rad\ m^{-2}$. The right panel is composed of four subpanels. From top to bottom, these show the PA versus time, the polarization profile, the Stokes $I$ dynamic spectrum, and the Stokes $V$ dynamic spectrum. In the profile panel, the gray, red, and blue curves denote Stokes $I$, linear polarization $L$, and circular polarization $V$, respectively. The gray shaded region indicates the burst window used for the polarization analysis. Relevant burst properties, including the burst time range, peak flux density and fluence, integrated and peak signal-to-noise ratio (SNR), polarization fractions, DM, and RM, are listed on its right side.}
    \label{fig:qu}
\end{figure*}

\clearpage
\begin{figure}
    \centering
    \includegraphics[width=0.47\linewidth]{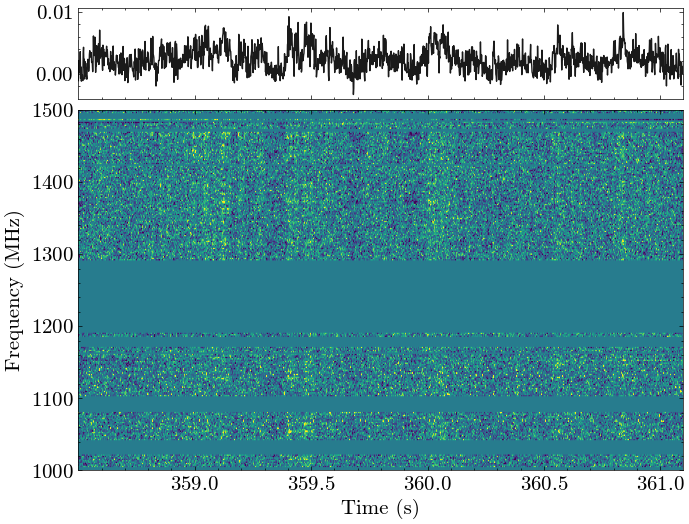}
    \includegraphics[width=0.47\linewidth]{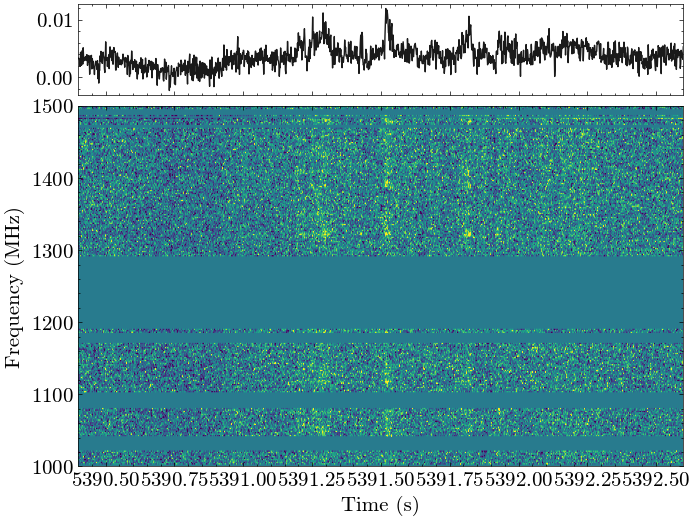}
    
    \includegraphics[width=0.48\linewidth]{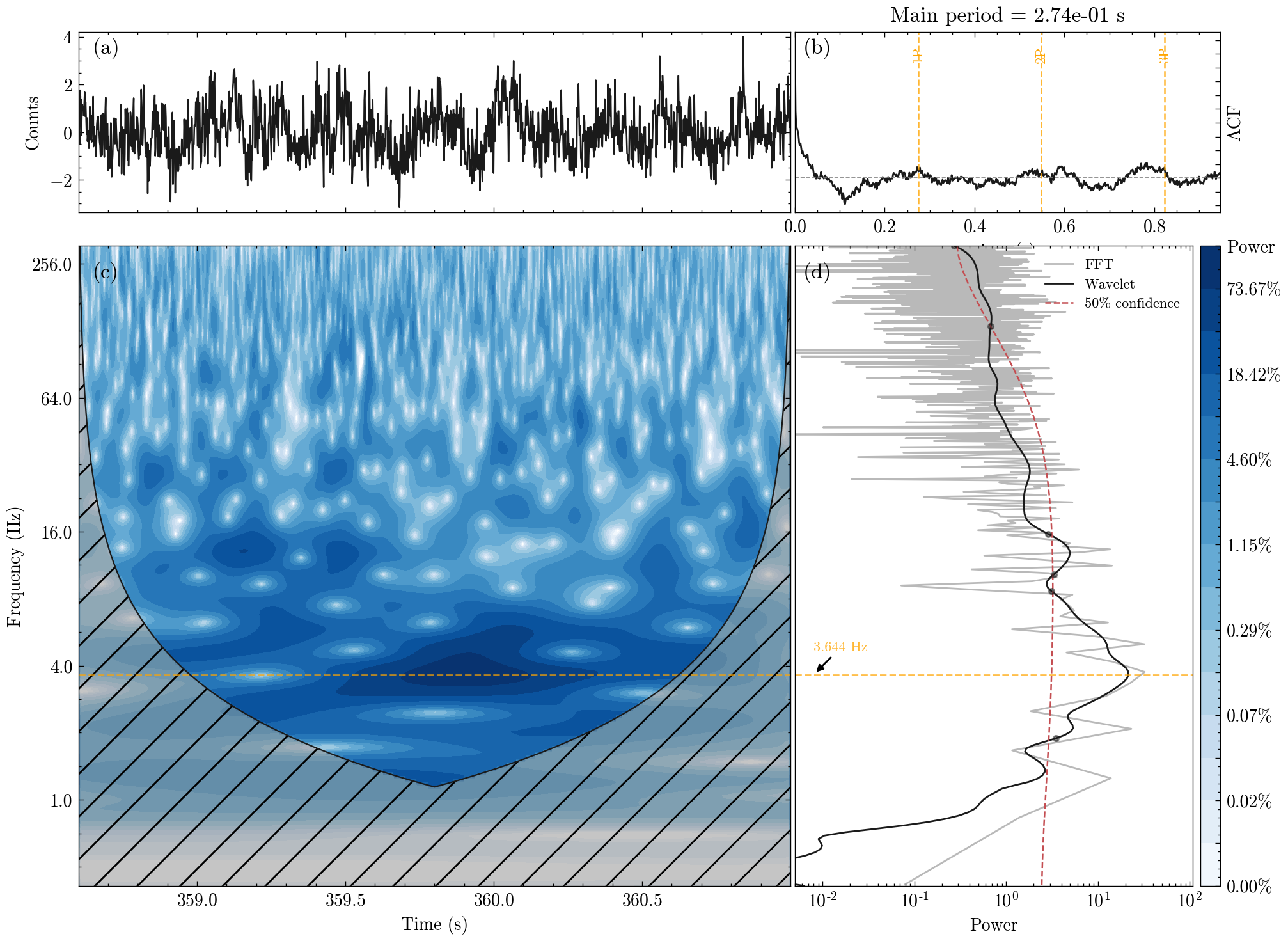}
    \includegraphics[width=0.48\linewidth]{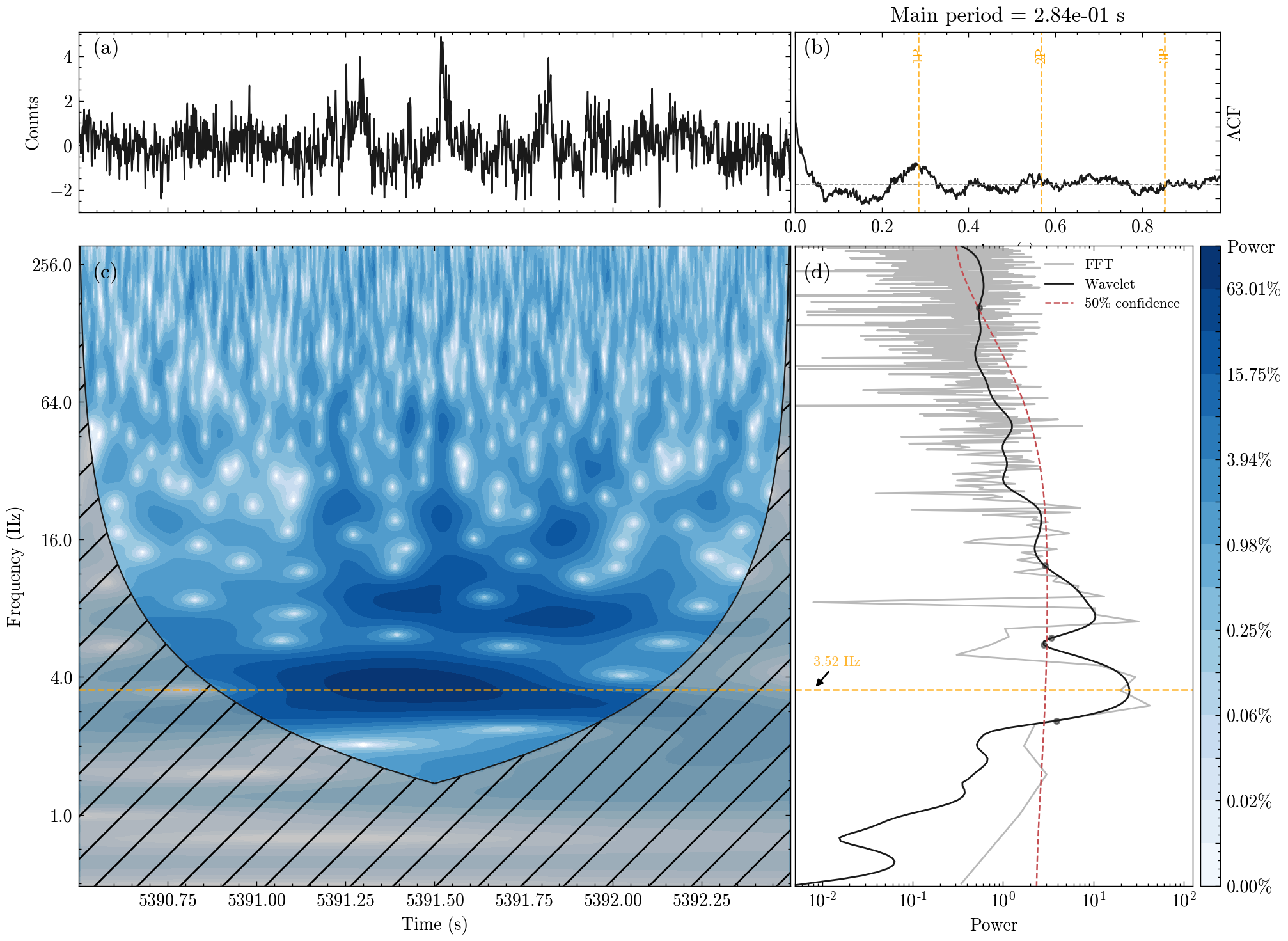}
    \caption{QPO-like behaviors were detected at 358.6-361.0 s and 5391-5392 s on 2026 February 6. For each panel: (a) the Stokes I light curve averaged over good channels; (b) the auto-correlation function (ACF), with the fundamental period (1P) and its harmonics (2P and 3P) marked by yellow dashed lines; (c) the wavelet power spectrum; and (d) the FFT spectrum together with the wavelet power spectrum averaged over time. A common periodicity of 3.644 Hz (0.274 s) is detected in both burst cycles. }
    \label{fig:qpo}
\end{figure}
\end{document}